\documentclass[a4paper,reqno,twoside,psamsfonts]{amsart}
\usepackage[latin1]{inputenc}
\usepackage[T1]{fontenc}
\usepackage{textcomp}
\usepackage[british]{babel}
\usepackage{pstricks}
\usepackage{newcent} 
\usepackage[dvips,pagebackref]{hyperref}
\usepackage{AUS-math}
\usepackage{bbold} 
\newcommand{\OA}{\emox{\mathcal{A}}}
\newcommand{\SC}{\emox{\overline{S}}}
\newcommand{\bbone}{\emox{\mathbb{1}}}
\newcommand{\XX}{\emox{\mathbb{X}}}
\providecommand{\href}[2]{#2}

\providecommand{\hypersetup}[1]{}
\begin{document}
\selectlanguage{british}
%
%
\title[Mathematics of the Quantum Zeno Effect]{Mathematics of the
Quantum Zeno Effect}
\author[A.\ U.\ Schmidt]{Andreas U.\ Schmidt}
\date{21st July 2003, revised 5th October 2004 including corrections from the published corrigendum
to~[64].}
\curraddr{Fraunhofer -- Institute for Secure Information Technology\\
Dolivostraße 15\\
64293 Darmstadt\\
Germany}
\address{Fachbereich Mathematik\\
  Johann Wolfgang Goethe-Universität\\
  60054 Frankfurt am Main, Germany}
  \email{\href{mailto:aschmidt@math.uni-frankfurt.de}{aschmidt@math.uni-frankfurt.de}}
  \urladdr{\href{http://www.math.uni-frankfurt.de/~aschmidt}{http://www.math.uni-frankfurt.de/\~{}aschmidt}}
\subjclass{\href{http://www.ams.org/msc/46Lxx.html}{46L60}, 
\href{http://www.ams.org/msc/46Lxx.html}{47D03}, 
\href{http://www.ams.org/msc/81Pxx.html}{81P15}, 
\href{http://www.ams.org/msc/81Rxx.html}{81R15}, 
\href{http://www.ams.org/msc/82Bxx.html}{82B10}} 
\thanks{\textit{\href{http://www.aip.org/pacs/}{PACS} 
classification.} 03.65.Xp, 03.65Db, 05.30.--d, 02.30.Tb}
\thanks{The author thanks Máté Matolcsi, Roman Shvidkoy,
as well as Harald Atmanspacher, Werner Ehm, and Tilmann Gneiting
for making their recent works available to him}
\begin{abstract}
 We present an overview of the mathematics underlying the quantum Zeno
effect. Classical, functional analytic results are put
into perspective and compared with more recent ones. This yields some new insights
into mathematical preconditions entailing the Zeno paradox, in particular
a simplified proof of Misra's and Sudarshan's theorem. We emphasise the complex-analytic
structures associated to the issue of existence of the Zeno dynamics. 
On grounds of the assembled material, we reason about possible future mathematical 
developments pertaining to the Zeno paradox and its counterpart, the anti--Zeno paradox,
both of which seem to be close to complete characterisations.
\end{abstract}
\maketitle
%
%
\section{Introduction}
The Zeno effect consists in the impediment of a quantum system's evolution
by frequent measurements performed on it. 
Apart from the generic quantum phenomenon of entanglement, 
it is probably the most striking difference
separating the classical from the quantum world, 
and an example for the sometimes
counterintuitive features of the latter.
In the most concise manifestation of the Zeno effect, a decaying
state of a quantum system, say an excited state of an atom, is 
conserved and prevented from decay simply by `looking at it'\ie
observing the presence of the undecayed state. That this observation
can clearly, in the quantum formalism, also be done by `doing nothing'
through a negative-result experiment \textit{not} observing the
decay products, see~\cite{KS02}, adds the same scent of magic (or `spookiness') 
to this effect that adheres to the Einstein--Podolsky--Rosen paradox.
The effect's name was aptly coined after the classical argument of Zeno 
that was meant to prove the impossibility of any real motion
(`a watched arrow never flies').

It is no wonder that such a baffling phenomenon has by now also entered popular 
science texts~\cite{b:WIC95}. The interest of the mathematical, theoretical, and
experimental physics communities in the effect --- which was formerly
seen as a mere curiosity, even possibly due only to a `wrong' interpretation
of quantum theory --- was rekindled by the seminal work of Misra and
Sudarshan~\cite{MS77}, which also
drew the present author into the subject, 
and inspired his recent works~\cite{AUS02A,AUS03}.

In the present survey, we report on the mathematical side of the story,
and
 put classical and more recent results into perspective.
More specifically, we concern ourselves with the stricter version
of the effect presented by the \textbf{Zeno paradox}~\cite{MS77}.
In its mathematical formulation using the projection postulate of the
standard interpretation of quantum mechanics to model the single
measurements by a projection $E$, the question that arises is
that of the existence of the limit
\[
W(t)=\lim_{n\to\infty}\bigl[EU(t/n)E\bigr]^n,
\]
where $U(t)$ is the original quantum dynamics --- which need in fact
\textit{not} necessarily be unitary as suggested by the use of the letter $U$.
That is, does this limit exist in an 
appropriate
topology on the
operators on the quantum mechanical Hilbert space of the system?
If so, is it a `sensible' quantum evolution by itself\ie is it continuous
and satisfies a (semi)group law? In the paradigmatic case where
$E$ is a rank one projection onto a single, decaying, initial state
$\psi_i$, written $|\psi_i\rangle\langle\psi_i|$ in Dirac's notation,
the paradox amounts to a complete impediment of the decay, if
$\psi_i$ is observed with infinite frequency.
Therefore, the Zeno paradox has been nicknamed `a watched pot never boils'
effect or `watchdog' effect by some~\cite{WHI00}.

This extreme manifestation of Zeno's effect is of conceptual
interest, although it is not a proper physical phenomenon since the
limit in question is not attainable due to the finite duration
of real world measurements, respectively, interactions involving a
\textit{finite} amount of energy. These principal upper bounds on
the frequency of measurements that can be exerted on a quantum system
are essentially a consequence of the time/energy uncertainty 
principle~\cite{PN94,PAT96,HNNPR98}. Nevertheless, the paradox is 
interesting both from a mathematical as well as from a physical 
viewpoint, for not only is the occurrence of the paradox, or rather sufficient 
conditions for it, also indicative for the Zeno effect at finite 
measurement frequency. But it is also very helpful for the study of the
\textbf{Zeno subspace}\ie the subspace of the full Hilbert space of the 
system to which the dynamics is confined in the \textbf{Zeno limit}
of infinitely frequent measurement. The general picture that has evolved
over the past few years~\cite{FPSS01,FP02,FP03,AUS02A,AUS03}, is that 
the emergent \textbf{Zeno dynamics} on the Zeno subspace is (more or less,
depending on the model considered) a free quantum dynamics, amended by 
specific boundary conditions. These interesting, and sometimes deep,
physical structures are in fact best studied in the firm framework
provided by the Zeno limit.

The present survey addresses physicists who might not be aware of
the functional analytic structures underlying Zeno's paradox,
as well as mathematicians who might not know of this special physical
application of these structures. We hope to incite the interest
of the two cultures in the subject, and to promote the mutual transfer of knowledge
about it. The main sources of the paper are 
the classical results by Kato~\cite{KAT78}, 
and Misra and Sudarshan~\cite{MS77}
which are put into perspective with the recent ones by
Matolcsi and Shvidkoy~\cite{MS01,MAS02,MAT03}, 
Exner and Ichinose~\cite{EI03}, 
Atmanspacher \textit{et al.}~\cite{AEG03},
and of the present author~\cite{AUS02A,AUS03}.

Yet, many interesting ramifications of  Zeno effect and paradox
are neglected, in particular,
we do not delve into the vast history of the subject, which dates
back to the 1930ies and is associated with the names of Turing and von Neumann.
The reader interested in this part of the story, as well as in the relation
of Zeno's effect to the controversies surrounding the interpretation of quantum theory,
is deferred to~\cite{HW86,NNP96,HW98,WHI00,GUS02}, and their extensive lists
of references.
Recent experiments confirming the Zeno effect~\cite{WBT01}, and its counterpart
the anti-Zeno effect~\cite{FGMR01} (see Section~\ref{sec:Z-AZ}) are
exciting, but also left out, as well as the possible practical applications
of the effects, and many theoretical treatments of
model cases, some of which also yield proposals for 
experiments~\cite{AER96,FKPS00,FP00,CAM01,TPN01,FHKPR02,FP02A,AUS03,NTY03}.
A subject which would also be relevant to the mathematical side of the matter,
but does only receive marginal consideration in Section~\ref{sec:gen-equi},
is the rather different manifestation of the Zeno effect caused by a
\emph{continuous} measurement performed on a system coupled to a
measurement device with a coupling strength approaching infinity.
Mathematical models for this, and the question of equivalence
of the different realisations of the Zeno paradox, have been 
treated elsewhere~\cite{BJ93,FP01,FP01A,FP02,FP03}.
Finally, counterexamples\ie the rather exceptional cases in which the
Zeno paradox does not emerge~\cite{AU97,MS01,MAS02}, are not expanded on.

The outline of the paper is as follows. In Section~\ref{sec:math} we gather
mathematical results which are fundamental for the examination of the Zeno
paradox, in the sense that they do not depend on specific information about 
the physical system considered, respectively, its mathematical model.
These results appear in the form of convergence theorems in the case
of semigroups in Section~\ref{sec:semigroup}, 
and existence theorems for the Zeno dynamics in the
quantum mechanical case of unitary groups treated by Misra's and
Sudarshan's Theorem. We show in Section~\ref{sec:unitary} 
how the proof of the latter can
be considerably shortened by using the former results. 
As a third case we present in Section~\ref{sec:modular}
our corresponding, 
recent result in the framework of modular flows of von Neumann algebras.
We highlight the crucial differences between the former and the latter two 
cases, hinging on analyticity domains.
This points to a possible way to achieve sharper characterisations 
of the Zeno paradox by use of  Payley--Wiener type arguments,
for which some preliminary thoughts are sketched in Section~\ref{sec:conc1}.

Section~\ref{sec:Conditions} presents a collection of more specific
conditions for the occurrence of the Zeno paradox, and results about
its physical consequences, formulated in operator theoretical frameworks.
The \textit{asymptotic Zeno condition} introduced in~\cite{AUS03},
and which is treated in Section~\ref{sec:AZC}, is efficient in that
it can be easily tested in concrete cases and enables the use of perturbation
theory for that purpose. This condition yields a mathematical formulation
of the Zeno paradox in the operator algebraic framework of quantum statistical
mechanics~\cite{b:BR79/81}, presented in Section~\ref{sec:AZC-app}.
As a consequence, we obtain the proper paradigmatic manifestation 
of the Zeno effect in quantum statistical mechanics --- the prevention
of return to equilibrium.
We further show in Section~\ref{sec:gen-equi}, that in many benign cases
it is possible to identify the Zeno subspace as well as the generator
of the Zeno dynamics acting on it. This enables the construction
of an important subclass of equilibrium states for the Zeno dynamics.
Note that examples for the Zeno effect in quantum statistical mechanics, more
specifically quantum spin systems, and the XY-model of a one-dimensional 
spin chain, are also treated in~\cite{AUS03}, but not reproduced here.
Finally in Section~\ref{sec:EI}, 
we contrast our own results with recent ones by Exner and Ichinose, 
which give a very sharp, abstract condition on the generator of the
Zeno dynamics that ensures existence of the Zeno limit.

Finally, Section~\ref{sec:phys} complements the foregoing abstract
presentation with some more physically-minded considerations. In particular,
we explain in Section~\ref{sec:geometry} how the reasons for the occurrence
of the Zeno effect and the Zeno paradox can be traced back to fundamental 
properties of the Hilbert space and its geometry, which entail the quadratic
short-time behaviour of quantum probabilities. On a quite different note,
and in opposition to the genericity of the quantum Zeno phenomenon 
stressed throughout the previous sections, Section~\ref{sec:Z-AZ}
presents the existence of the opposite phenomenon --- the \textit{anti-Zeno}
effect. Under certain conditions, there can be a region of measurement 
frequencies at which the quantum evolution is spurred rather than impeded.
In the extreme case, even an anti--Zeno \emph{paradox}\ie
induction of instantaneous decay, seems conceivable. We present
two remarkable, recent results about necessary \emph{and} sufficient 
conditions for Zeno and anti--Zeno paradox, based on the asymptotics of the
state's
energy distribution, in Section~\ref{sec:EDF}.

Some conclusions and open questions are noted in Section~\ref{sec:conc}.
\section{Mathematical Foundations}
\label{sec:math}
\subsection{Analytic Semigroups}
\label{sec:semigroup}
One important case in which the Zeno dynamics is sure to exist presents
itself within the theory of analytic semigroups. Many fundamental results
in that field are concerned with \textbf{Trotter's product formula}
\[
\ee^{t (A+B)} = \lim_{n\to\infty} \Bigl[ \ee^{\frac{t}{n}A}\ee^{\frac{t}{n}B}\Bigr]^n,
\]
which holds\eg if $A$ is the generator of a contractive semigroup and $B$ a
dissipative operator on a Banach space 
$\mathcal{B}$, see~\cite[Corollary~3.1.31]{b:BR79/81},
and~\cite{b:CHE74} for an extensive treatment of product formulae.
If $B$ is replaced by a projection 
$E$, one speaks of a \textbf{degenerate
product formula} of the form
\[
S(t)=\lim_{n\to\infty} \Bigl[ \ee^{\frac{t}{n}A} E \Bigr]^n,
\]
and wants to determine conditions under which this limit exists and defines a
strongly continuous semigroup on the invariant subspace 
$S(0)\mathcal{B}\subset E\mathcal{B}$ (note that 
$S(0)$ is a projection in this case),
a so called \textbf{degenerate semigroup}.

The one result in that vein which we want to present here is based on
Kato's work on non-densely defined sesquilinear forms~\cite{KAT78}. It
says, essentially, that the Zeno dynamics \textit{always} exists for
semigroups which are analytic in a sector in \CC with positive opening
angle.
\begin{theo}[{\cite[Theorem~4]{AU97}}]\labelT{analytic-semigroups}
Let $-A$ be the generator of a semigroup $(\ee^{-zA})$, $z\in\Sigma(\theta)$
on a Hilbert space \HS,  which is holomorphic in an open sector 
$\Sigma(\theta)=\bigl\{z\in\CC\bigm|z\neq0, \ABS{\arg z}<\theta\bigr\}$ for
some $\theta\in(0,\pi/2]$. Assume $\NORM{\smash[t]{\ee^{-zA}}}\leq 1$ for all
$z\in\Sigma(\theta)$, and let $E$ be an orthogonal projection.
Then
\[
S(t)\phi=\lim_{n\to\infty} \Bigl[ \ee^{-\frac{t}{n}A} E \Bigr]^n\phi
\]
exists for all $\phi\in\HS$, $t\geq0$, and defines a degenerate semigroup
$(S(t))_{t\geq0}$.
\end{theo}
The generality of this result is quite remarkable. It makes no assumptions on the 
projection and also does not depend on other technical details, for instance
whether or not \HS is separable. The strength of the assumption lies exclusively
in the analyticity domain. What this result already shows is that the Zeno
effect is by no means restricted to \textit{unitary} evolutions.
The following lucid explanation of its proof is taken almost literally from~\cite{AU97},
see also~\cite{MAS02}.

Let $a\colon D(a)× D(a)\to \CC$ be a sesquilinear form with
a subspace $D(a)\subset\HS$ as domain. Assume that $a$ is \textbf{semibounded}\ie
\[
\exists \lambda\in\RR\colon\forall0\neq\phi\in D(a)\colon\quad
\NORM{\phi}_a^2\DEF\RE a\IPROD{\phi}{\phi}+\lambda\IPROD{\phi}{\phi}_\HS>0,
\]
where $\IPROD{\cdot}{\cdot}_\HS$ is the scalar product of \HS, 
and moreover that it is
\textbf{sectorial}\ie
\[
\exists M>0\colon\quad
\ABS{\IM a\IPROD{\phi}{\phi}} \leq
M \bigl(\RE a\IPROD{\phi}{\phi} +\lambda\IPROD{\phi}{\phi}_\HS\bigr),
\]
and \textbf{closed}\ie $(D(a),\NORM{\cdot}_a)$ is complete. On the closure 
$K=\overline{D(a)}$, define the operator $A$ associated with $a$ by
\begin{align*}
 D(A) &= \bigl\{ \phi\in D(a)\bigm|
   \exists\psi\in K\colon\ a\IPROD{\phi}{\chi}=\IPROD{\psi}{\chi}_\HS\
   \forall\chi\in D(a)\bigr\},\\
  A \phi &= \psi.
\end{align*}
Then $-A$ generates a $C_0$-semigroup $(\ee^{-tA})_{t\geq0}$ on $K$.
If $P(K)$ denotes the orthogonal projection onto $K$, then we can define
the operators 
$\ee^{-ta}$
on \HS by $\ee^{-tA}P(K)$, and they constitute a
degenerate semigroup on \HS. If $b$ is a second
semibounded, closed, sesquilinear form, then their sum $a+b$ defined by
\begin{align*}
  D(a+b) &= D(a)\cap D(b),\\
 (a+b)\IPROD{\phi}{\psi} &= a\IPROD{\phi}{\psi} + b\IPROD{\phi}{\psi},
\end{align*}
is again semibounded and closed. In this case holds the
following product formula.
\begin{theo}[{\cite[Theorem and Addendum]{KAT78}}]
  For $\phi\in\HS$ holds
\[
\ee^{-t(a+b)}\phi=\lim_{n\to\infty}
\Bigl[ \ee^{-\frac{t}{n}a}\ee^{-\frac{t}{n}b}\Bigr]^n\phi
\]
for all $t>0$.
\end{theo}
To make contact with \refT{analytic-semigroups}, we note that
under the conditions stated there, there exists a closed, semibounded,
sesquilinear form $a$ associated with the generator 
$A$, see~\cite[Section VI.2.1, Theorem~2.7]{b:KAT76}. 
In particular, this form is sectorial in the sense above, by
Theorem 1.2 of~\cite{ABH01}.
On the other hand,
the projection $E$ defines the form $b$ by
\begin{align*}
  D(b)&= E\HS,\\
  b\IPROD{\phi}{\psi}&\equiv 0\quad\text{on }E\HS,
\end{align*}
entailing $\ee^{-tb}=E$ for all $t\geq0$. Thus
\[
S(t)\phi=\ee^{-t(a+b)}\phi=\lim_{n\to\infty}
\Bigl[ \ee^{-\frac{t}{n}a}E\Bigr]^n\phi, 
\quad \phi\in\HS,
\]
is the semigroup of the conclusion of \refT{analytic-semigroups}.

Well, on this level the whole issue of the Zeno paradox might seem resolved.
But this is not quite the case, as is shown by the counterexamples 
constructed in~\cite{AU97} in the broader context
of positive semigroups on Banach spaces,
and in~\cite{MS01,MAS02} even for unitary semigroups on Hilbert spaces. 
That, and why, things are a bit more involved in the Hilbert space 
case, will be seen in the following.
\subsection{Unitary Groups}
\label{sec:unitary}
It is not without irony that the above fundamental results on
degenerate semigroups appeared in~\cite{KAT78},
one year after the publication of the seminal article of Misra and 
Sudarshan~\cite{MS77} --- one of the main sources of inspiration
for everyone who is today interested in the Zeno effect.
For, as we will show now, the analytic semigroup results
can be used to simplify the original proof of
Misra's and Sudarshan's existence theorem for Zeno dynamics
significantly. 
\begin{Ltheo}{unitary}
  Let $U(t)=\ee^{\ii tH}$ be a unitary group of operators on a
Hilbert space \HS, with nonnegative, self-adjoint generator $H$, and
$E$ an orthogonal projection. 
Assume that the limits
\[
T(t)=\slim_{n\to\infty}\bigl[EU(t/n)E\bigr]^n
\]
exist for all $t\in\RR$, are weakly continuous in $t$, 
and satisfy the initial condition
\[
\wlim_{t\to 0}T(t)=E.
\] 
Then, $T(t)$ is a degenerate group of unitaries on $E\HS$.
\end{Ltheo}
\begin{proof}
  Since $H$ is nonnegative, $U(t)$ extends to a holomorphic, operator-valued
function $U(z)$ in the upper half-plane $\mathbb{H}_+=\bigl\{z\bigm|\IM z>0\bigr\}$ 
with boundary value $U(t)$. \refT{analytic-semigroups} entails the existence
of $T(z)=\lim_{n\to\infty}[EU(z/n)E]^n$ on every ray
$\bigl\{z=r\ee^{\ii\theta}\bigm| \theta\in(0,\pi), r>0\bigr\}$, and
therefore in whole $\mathbb{H}_+$. In particular, $(W(\ii s))_{s\geq0}$ is
a degenerate, analytic semigroup, and long known structure theorems
for these~\cite{b:HP65} ensure the existence of a positive operator 
$B$ and a projection $G$ such that $W(\ii s)=\ee^{-sB}G=G\ee^{-sB}$.
Thus, for all $\phi\in\HS$ holds $W(\ii s)\phi=G\ee^{-sB}G\phi$, and
the identity theorem for vector-valued, holomorphic 
functions~\cite[Theorem~3.11.15]{b:HP65} ensures the identity
$W(z)\phi=G\ee^{\ii zB}G\phi$ for all $z\in\mathbb{H}_+$. In turn,
this yields the semigroup property $W(z_1+z_2)=W(z_1)W(z_2)$ for
all $z_1$, $z_2\in\mathbb{H}_+$, since $\ee^{\ii zB}$ commutes with $G$.
We obtain the integral representation
\begin{alignat*}{2}
W(z)&=\slim_{n\to\infty}\bigl[EU(z/n)E\bigr]^n 
& & \\
&=\slim_{n\to\infty}
\frac{(z+\ii)^2}{2\pi\ii}\int_{-\infty}^\infty
\frac{\bigl[EU(t/n)E\bigr]^n}{(t+\ii)^2(t-z)}\dd t 
& & \ \left(\text{\small Cauchy's formula}\right)\\
&=
\frac{(z+\ii)^2}{2\pi\ii}\int_{-\infty}^\infty
\frac{\slim\limits_{n\to\infty}\bigl[EU(t/n)E\bigr]^n}{(t+\ii)^2(t-z)}\dd t
& &\ \left(\text{\parbox{3.15cm}{\small uniform boundedness\\
 of the integrand}}\right)\\
&=\frac{(z+\ii)^2}{2\pi\ii}\int_{-\infty}^\infty
\frac{T(t)}{(t+\ii)^2(t-z)}\dd t,
& &\ \left(\text{\small convergence assumption}\right)\\
\end{alignat*}
for $\IM z>0$. On the other hand
\[
0=\frac{(z+\ii)^2}{2\pi\ii}\int_{-\infty}^\infty
\frac{T(t)}{(t+\ii)^2(t-z)}\dd t,
\]
for $\IM z<0$, both of which we now use to carry the semigroup 
property forward from the complex 
domain $\mathbb{H}_+$ to the boundary \RR.
Namely writing explicitly $z=s±\ii\eta$, $\eta>0$, and 
adding the two integral representations above yields
\[
\IPROD{\phi}{W(s+\ii\eta)\psi}=
\frac{(s+\ii+\ii\eta)^2}{\pi}
\int_\RR \frac{\IPROD{\phi}{T(t)\psi}}{(t+\ii)^2}
\frac{\eta}{(t-s)^2+\eta^2}\dd t,
\]
for any two vectors $\phi$, $\psi\in\HS$.
The right hand side is nothing but the Poisson transformation of 
the function $\IPROD{\phi}{T(s)\psi}$, 
modified by the convergence factor
$(t+\ii)^{-2}$,  and it reproduces
this function $s$-almost everywhere as $\eta\to 0_+$.
But, since $T(t)$ is weakly continuous, we obtain
\[
\wlim_{\eta\to 0_+} W(t+\ii\eta) = T(t),\quad\text{for all }t\in\RR.
\]
In turn, $T(t)=G\ee^{\ii tB}G$ for all $t$,
yielding in particular the strong continuity of $T(t)$. 
Thus $T(t)T(t)^\ast=G\ee^{\ii tB}\ee^{-\ii tB}G=G$ for all $t$, 
which together with the initial condition implies $G=E$. 
This finally shows $T(t)=E\ee^{\ii tB}E$ for all $t\in\RR$.
Since the explicit form of the limit in $n$ entails $T(-t)=T(t)^\ast$, we
conclude that $(T(t))_{t\in\RR}$ is a strongly continuous group
of operators which are unitary on $E\HS$,
and the proof is completed.
\end{proof}
\subsection{Modular Automorphisms of von Neumann Algebras}
\label{sec:modular}
The last, and technically most involved, abstract result on  Zeno dynamics
is about modular flows of von Neumann algebras. This might seem
of purely mathematical interest, were it not for the close connection
between these flows and the dynamical flows of systems in quantum statistical
mechanics, associated with KMS states, which in turn are the paradigm
for thermal equilibrium (we refer to~\cite{b:BR79/81} for the background).
But also from our present viewpoint it is interesting to 
compare the case of modular flows with the other two presented above,
since here the generator of the original dynamics is not semibounded.
It rather is such that the negative spectral degrees of freedom are
exponentially damped in the following sense, cf.~\cite[Section~V.2.1]{b:HAA92}.
To fix notation, we let \OA be a von Neumann algebra with faithful, normal state
$\omega$, represented on the Hilbert space 
$\HS=\overline{\OA\Omega}$ 
with cyclic and separating vector $\Omega$ associated with $\omega$.
Let $\Delta$ be the modular operator of $(\OA,\Omega)$, generating the
modular flow $U(t)=\Delta^{\ii t}$. Write
$\Delta=\ee^{-K}$, $U(t)=\ee^{-\ii t K}$. Let $E^{(-)}_\kappa$, $\kappa>0$,
be the spectral projection of $K$ for the interval $[-\infty,-\kappa]$.
Then holds
\[
\NORM{\smash[t]{E^{(-)}_\kappa} A\Omega}\leq\ee^{-\kappa/2}\NORM{A},
\quad\text{for all }A\in\OA,
\]
and therefore the vectors in $\OA\Omega$ are in the domain of $\Delta^\alpha$
for $0\leq\alpha\leq1/2$.

This difference to the quantum mechanical case entails that the
original evolution is no longer analytically extensible to the
whole upper half-plane, but only to a strip of positive width
(which lies, due to a notorious change of sign, in the \textit{lower} half-plane).
For us this means in particular that the analyticity domain contains 
no open sector, and we cannot use \refT{analytic-semigroups} 
to infer the existence of the Zeno dynamics within it. Therefore,
we have to use the independent, original strategy of~\cite{MS77}
and adapt it accordingly. A second, more technical point is that
we now have to handle the generically unbounded operators in 
question with some additional care.
For the reader's convenience, 
and to ease comparison with Section~\ref{sec:unitary},
we reproduce the results and proofs of~\cite{AUS02A} rather completely.
The tenets of their application to quantum statistical mechanics are 
described in Section~\ref{sec:AZC-app} below.
\begin{Ltheo}{modular}
Let $E\in\OA$ be 
a projection. 
Set $\OA_E\DEF E\OA E$, and define a subspace of \HS by 
$\HS_E\DEF\overline{\OA_E\Omega}\subset E\HS$.
Assume:
  \begin{enumerate}
  \item\label{ass1}
For all $t\in\RR$, the strong operator limits 
\[
W(t)\DEF\slim_{n\to\infty}
\bigl[E \Delta^{\ii t/n} E\bigr]^n
\]
exist, are weakly continuous in $t$, and satisfy the 
initial condition 
\[
\wlim_{t\to0}W(t)=E.
\]
  \item For all $t\in\RR$, the following limits exist:
\[
W(t-\ii/2)\DEF\slim_{n\to\infty}
\bigl[E \Delta^{
\ii
 (t-\ii/2)/n} E\bigr]^n,
\]
where the convergence is strong on the common, dense
domain $\OA\Omega$.
\end{enumerate}
Then the $W(t)$ form a strongly continuous group of unitary operators
on  $\HS_E$. The group $W(t)$
induces an automorphism group $\tau^E$ of  
$\OA_E$ by
\[
\tau^E\colon\OA_E\ni A_E\longmapsto \tau^E_t(A_E)\DEF W(t)A_EW(-t)=
W(t)A_EW(t)^\ast,
\]
such that $(\OA_E,\tau^E)$ is a 
$W^\ast$-dynamical system. The vectors 
$W(z)A_E\Omega$,  $A_E\in\OA_E$, are holomorphic in the strip $0<-\IM{z}<1/2$ and 
continuous on its boundary.
\end{Ltheo}
Notice that $\OA_E$ is a von Neumann subalgebra of 
$\OA$, see~\cite[Corollary~5.5.7]{b:KR83/86}, for which  $\Omega$ is
cyclic for $\HS_E$, and separating. Thus, $\Omega$ induces a faithful
representation of $\OA_E$ on the closed
Hilbert subspace $\HS_E$, and thus all notions above are well-defined.
The remainder of this section contains the proof
of the above theorem, split into several lemmas. 
Set $S\DEF\bigl\{z\in\CC\bigm|-1/2<\IM z<0\bigr\}$.
Define operator-valued functions
\[
F_n(z)\DEF \bigl[E \Delta^{\ii z/n} E\bigr]^n, \quad
\text{for }z\in\SC,\ n\in\NN.
\]
The $F_n(z)$ are operators whose domains of definition contain
the common, dense domain $\OA\Omega$. They depend holomorphically on $z$ 
in the sense that the vector-valued functions $F_n(z)A\Omega$ are 
holomorphic on $S$ and continuous on \SC for every $A\in\OA$. For this
and the following lemma see~\cite[Section~2.5, Section~5.3, 
and Theorem~5.4.4]{b:BR79/81}.
\begin{Llemm}{FnBound}
  For $z\in\SC$ and $\psi\in D(\Delta^{\ABS{\IM z}})$ holds the estimate
\[
\NORM{F_n(z)\psi}\leq\NORM{\psi},
\]
for all $n\in\NN$.
\end{Llemm}
\begin{proof} 
Define vector-valued functions 
$
f_k^{\psi,n}(z)\DEF \bigl[E\Delta^{\ii z/n}E\bigr]^k \psi.
$
These are well-defined for $z\in\SC$, $\psi\in D(\Delta^{\ABS{\IM z}})$ and
all $k\leq n$, since for such $\psi$, $z$ we have
$\bigl[E\Delta^{\ii z/n}E\bigr]^{k-1}\in D(E\Delta^{\ii z/n}E)$.
Approximate $f_{k-1}^{\psi,n}(z)$ by elements of the form $A_l\Omega$, $A_l\in\OA$.
Then for any $B\in\OA$ holds
\begin{align*}
  \ABS{\SPROD{\smash{B\Omega}}{\smash{E\Delta^{\ii z/n}E A_l  \Omega}}} &=
  \ABS{\SPROD{\Omega}{\smash{B^\ast E \Delta^{\ii z/n}E A_l 
    \Delta^{-\ii z/n}\Omega}} } \\ &
 =\ABS{\omega(B^\ast E\sigma_{z/n}(E A_l)) }\\ 
&\leq \NORM{B^\ast E\Omega}\NORM{EA_l\Omega} \\ &\leq
  \NORM{B}\NORM{A_l\Omega}.
\end{align*}
Here, $\omega$ is the state on \OA associated with the cyclic and separating
vector $\Omega$ (we always identify elements of \OA with their representations
on \HS), and $\sigma$ denotes the modular group. The first estimate above follows explicitly
from the corresponding property of $\sigma$, see~\cite[Proposition~5.3.7]{b:BR79/81}
(the connection between faithful states of von Neumann algebras and
KMS states given by Takesaki's Theorem~\cite[Theorem~5.3.10]{b:BR79/81} is 
used here and in the following).
This means $\NORM{E\Delta^{\ii z/n}E A_l\Omega}\leq\NORM{A_l\Omega}$, and since 
$A_l\Omega\longrightarrow f_{k-1}^{\psi,n}(z)$ in the norm of \HS, it follows
$\NORM{\smash[t]{f_k^{\psi,n}(z)}}\leq\NORM{\smash[t]{f_{k-1}^{\psi,n}(z)}}$. 
Since this holds for all $k=1,\ldots,n$, we see
\[
\NORM{F_n(z)\psi} = \NORM{\smash[t]{f_n^{\psi,n}(z)} } 
\leq \ldots \leq \NORM{\smash[t]{f_1^{\psi,n}(z)} }\leq \NORM{\psi},
\]
as desired.
\end{proof}
The estimate proved above also yields that the $F_n$ are closeable.
We will denote their closures by the same symbols in the following.
\begin{Llemm}{FnInt}
 For $z\in S$ holds the representation
\begin{equation}\labelE{FnInt}
F_n(z)A\Omega =
\frac{(z+\ii )^2}{2\pi\ii}\int_{-\infty}^{\infty}
\frac{F_n(t-\ii /2)A\Omega}{(t+\ii /2)^2(t-\ii /2-z)}-
\frac{F_n(t)A\Omega}{(t+\ii )^2(t-z)}\;\dd t.
\end{equation}
where the integrals are taken in the sense of Bochner. One further has
\begin{equation}\labelE{Fn0}
0 = \frac{1}{2\pi\ii}\int_{-\infty}^{\infty}
\frac{F_n(t-\ii /2)A\Omega}{(t+\ii /2)^2(t-\ii /2-z)}-
\frac{F_n(t)A\Omega}{(t+\ii )^2(t-z)}\;\dd t,  
\end{equation}
for $z\not\in\SC$.
\end{Llemm}
\begin{proof}
  By Cauchy's theorem for vector-valued functions~\cite[Theorem~3.11.3]{b:HP65},
we can write
\[
\frac{F_n(z)A\Omega}{(z+\ii )^2}=
\frac{1}{2\pi\ii}\oint\frac{F_n(\zeta)A\Omega}{(\zeta+\ii )^2(\zeta-z)}\;\dd\zeta,
\]
where the integral runs over a closed, positively oriented contour in $S$, 
which encloses $z$. We choose this contour to be the boundary of the rectangle
determined by the points 
$\bigl\{R-\ii \EPS,-R-\ii \EPS,-R-\ii (1/2-\EPS),R-\ii (1/2-\EPS)\bigr\}$ 
for $R>0$, $1/4>\EPS>0$. By \refL{FnBound}, the norms of the integrals over 
the paths parallel to the real line stay bounded as $R\to\infty$, 
while those of the integrals parallel to the imaginary axis vanish. Thus
\begin{gather*}
\frac{F_n(z)A\Omega}{(z+\ii )^2} =
\frac{1}{2\pi\ii}\int_{-\infty}^{\infty}
\frac{F_n(t-\ii (1/2-\EPS))A\Omega}{(t+\ii (1/2+\EPS))^2(t-\ii (1/2-\EPS)-z)}
\\ -\frac{F_n(t)A\Omega}{(t+\ii (1-\EPS))^2(t-\ii \EPS-z)}\;\dd t.
\end{gather*}
For $0<\EPS_0<\min\{\ABS{\IM z},\ABS{1/2-\IM z}\}$ and all 
$\EPS$ such that $0\leq\EPS\leq\EPS_0$,  the integrand is bounded in norm by
$\NORM{A}\bigm/\bigl[(1+t^2)\min\{\ABS{\IM z - \EPS_0} , \ABS{\IM z - (1/2-\EPS_0)}\}\bigr]$. 
Since moreover, in the strong sense and pointwise in $t$, 
$\lim_{\EPS\to 0}F_n(t-\ii \EPS)A\Omega=F_n(t)A\Omega$, and 
$\lim_{\EPS\to 0}F_n(t-\ii (1/2-\EPS))A\Omega=F_n(t-\ii /2)A\Omega$,
 the conditions for the application of the vector-valued 
Lebesgue theorem on dominated convergence~\cite[Theorem~3.7.9]{b:HP65} are given,
and the desired representation follows in the limit $\EPS\to0$.
The vanishing of the second integral follows analogously.
\end{proof}
\begin{Llemm}{HoloLim}
  The strong limits $F(z)\DEF\slim_{n\to\infty}F_n(z)$, $z\in S$, 
are closeable operators with common, dense domain $\OA\Omega$ (we 
denote their closures by the same symbols). The integral representation
\begin{equation}\labelE{FInt}
F(z)A\Omega =
\frac{(z+\ii )^2}{2\pi\ii}\int_{-\infty}^{\infty}
\frac{W(t-\ii /2)A\Omega}{(t+\ii /2)^2(t-\ii /2-z)}-
\frac{W(t)A\Omega}{(t+\ii )^2(t-z)}\;\dd t
\end{equation}
holds good, and the functions $F(z)A\Omega$ are holomorphic on $S$, for all 
$A\in\OA\Omega$. There exists a projection $G$ and a positive operator $\Gamma$
such that $\Gamma=G\Gamma=\Gamma G$, and ${\Gamma}^{4iz}=F(z)$ for all
$z\in S$.
\end{Llemm}
\begin{proof}
  Using \refL{FnBound}, we see that the norm of the integrand in \eqref{eq:FnInt} 
is uniformly bounded in $n$ by 
$2\NORM{A}\bigm/\bigl[(1+t^2)\min\{\ABS{\IM z},1/2-\ABS{\IM z}\}\bigr]$,
which is integrable in $t$. Furthermore, $F_n(t)A\Omega$ and $F_n(t-\ii /2)A\Omega$ 
converge in norm to $W(t)A\Omega$ and $W(t-\ii /2)A\Omega$, respectively, by 
assumptions~i) and~ii) of \refT{modular}.
Thus, we can again apply Lebesgue's theorem on dominated convergence 
to infer the existence of the limits $\lim_{n\to\infty}F_n(z)A\Omega$
for all $A\in\OA$. This defines linear operators on the common, dense domain
$\OA\Omega$. Again, by the estimate of \refL{FnBound}, we have 
$F(z)A_n\Omega\to0$ if $\NORM{A_n}\to0$, and therefore the
$F(z)$ are closeable. The validity of \refE{FInt} is then clear.
Since the bound noted above is uniform in $n$, and all the functions
$F_n(z)A\Omega$ are holomorphic in $S$, we can apply the Stieltjes--Vitali
theorem~\cite[Theorem~3.14.1]{b:HP65} to deduce the stated holomorphy of $F(z)A\Omega$.
We now consider the operators $F(-\ii s)$, $0<s<1/2$. Using the same properties
of $\Delta$, $E$, one sees that these operators are self-adjoint, and in fact,
positive: Namely, the limits are densely defined, symmetric and closeable
operators, and an analytic vector for $\Delta^{1/2}$ is also analytic
for $F(-\ii s)$, $0<s<1/2$. Thus the $F(-\ii s)$ possess a common, dense
set of analytic vectors. Under these circumstances,  the $F(-\ii s)$ are 
essentially self-adjoint, and we denote their unique, self-adjoint extension
by the same symbol. 
We now follow~\cite{MS77} to show that the functional equation 
$F(-\ii (s+t))=F(-\ii s)F(-\ii t)$ holds for $s$, $t>0$ such that 
$s+t<1/2$. To this end, consider first the case that $s$ and $t$ are
rationally related\ie there exist $p$, $q\in\NN$ such that
\[
\frac{s+t}{r(p+q)}=\frac{s}{rp}=\frac{t}{rq}, \quad\text{for all }r\in\NN.
\]
Then
\[
\left[E\Delta^{\frac{s+t}{r(p+q)}}E\right]^{r(p+q)}A\Omega=
\left[E\Delta^{\frac{s}{rp}}E\right]^{rp}
\left[E\Delta^{\frac{t}{rq}}E\right]^{rq}A\Omega,\quad A\in\OA,
\]
from which the claim follows in the limit $r\to\infty$.
The general case follows since $F(-\ii s)A\Omega$ is holomorphic
and therefore also strongly continuous in $s$ for all $A\in\OA$.
Now set ${\Gamma}=F(-\ii /4)$. By the spectral 
calculus for unbounded operators~\cite[Section~5.6]{b:KR83/86}, 
the positive powers
$\Gamma^\sigma$ exist for $0<\sigma\leq1$, and are positive operators
with domain containing the common, dense domain $\OA\Omega$. They
satisfy the functional equation 
${\Gamma}^{\sigma+\tau}=
{\Gamma}^{\sigma}\CDOT{\Gamma}^{\tau}$
for $\sigma$, $\tau>0$ such that $\sigma+\tau\leq1$, and where
$\CDOT$ denotes the closure of the operator product. The solution
to this functional equation with initial condition 
${\Gamma}=F(-\ii /4)$ is unique and thus it follows
${\Gamma}^\sigma=F(-\ii \sigma/4)$, since the operators $F$ 
satisfy the same functional equation, and all operators in
question depend continuously on $\sigma$, in the strong sense
when applied to the common core $\OA\Omega$. For $1/4\leq s< 1/2$ 
we have $F(-\ii s)=F(-\ii /4)F(-\ii (s-1/4))={\Gamma}F(-\ii (s-1/4))=
{\Gamma}{\Gamma}^{4s-1}={\Gamma}^{4s}$,
which finally shows the identity $F(-\ii s)=\Gamma^{4s}$ for
$0<s<1/2$. Now, for every $A\in\OA$, 
${\Gamma}^{4iz}A\Omega$ extends to a holomorphic
function on $S$ which coincides with $F(z)A\Omega$ on the segment 
$\{-\ii s\mid 0<s<1/2\}$ as we have just seen. 
The identity theorem for vector-valued, 
holomorphic functions~\cite[Theorem~3.11.5]{b:HP65} then implies
${\Gamma}^{4iz}A\Omega=F(z)A\Omega$, $z\in S$ and all 
$A\in\OA$. Thus ${\Gamma}^{4iz}=F(z)$ holds on $S$ as an 
identity of densely defined, closed operators. Setting $G=P([0,\infty))$,
where $P$ is the spectral resolution of the identity for ${\Gamma}$,
we see that we can write ${\Gamma}=G\Gamma =\Gamma G$, concluding this
proof.
\end{proof}
\begin{Llemm}{Boundary}
  It holds $G=E$, and $W(t)=E\Gamma^{4it}E$,
for all $t$.
\end{Llemm}
\begin{proof}
Using~\eqref{eq:FInt} we can write, adding a zero contribution to that 
integral representation,
\begin{multline*}
  \SPROD{B\Omega}{F(t-\ii \eta)A\Omega} =\\
\frac{(t+\ii -\ii \eta)^2}{2\pi\ii}\int_{-\infty}^{\infty}\;\dd s\Biggl\{
\frac{\SPROD{B\Omega}{W(s-\ii /2)A\Omega}}{(s+\ii /2)^2(s-t-\ii /2+\ii \eta)}-
\frac{\SPROD{B\Omega}{W(s)A\Omega}}{(s+\ii )^2(s-t+\ii \eta)}-\\
-\frac{\SPROD{B\Omega}{W(s-\ii /2)A\Omega}}{(s+\ii /2)^2(s-t-\ii /2-\ii \eta)}+
\frac{\SPROD{B\Omega}{W(s)A\Omega}}{(s+\ii )^2(s-t-\ii \eta)}
\Biggr\},
\end{multline*}
where the integral over the last two terms is zero, as can be seen
from~\eqref{eq:Fn0} and the same arguments that were used to 
derive~\eqref{eq:FInt}. This yields
\[
= 
\frac{(t+\ii -\ii \eta)^2}{\pi}\int_{-\infty}^{\infty}
\frac{\eta\cdot\SPROD{B\Omega}{W(s-\ii /2)A\Omega}}{(s+\ii /2)^2((s-t-\ii /2)^2+\eta^2)}-
\frac{\eta\cdot\SPROD{B\Omega}{W(s)A\Omega}}{(s+\ii )^2((s-t)^2+\eta^2)}
\;\dd s.
\]
As $\eta\to0_+$, the first term under the integral vanishes,
while the second reproduces the integrable function
$\SPROD{A\Omega}{W(t)B\Omega}/(t+\ii )^2$ as the boundary
value of its Poisson transformation. Thus we have seen
\[
\lim_{\eta\to0_+}
\SPROD{A\Omega}{F(t-\ii \eta)B\Omega}=\SPROD{A\Omega}{W(t)B\Omega},
\]
for given $A$, $B\in\OA$, and almost all $t\in\RR$.
Since the integral is uniformly bounded in $\eta$,
the boundary value of this Poisson transformation
is continuous in $t$, see\eg~\cite[Section~5.4]{b:BRE65}.
The same holds for \SPROD{A\Omega}{W(t)B\Omega} by assumption~i)
of \refT{modular}, and therefore the limiting identity at $\eta=0$
follows for all $t$. 
On the other hand, since $G\Gamma^{4it}G$ is strongly continuous in $t$,
we have $\lim_{\eta\to0_+}
\SPROD{A\Omega}{F(t-\ii \eta)B\Omega}=
\SPROD{A\Omega}{G\Gamma^{4it}GB\Omega}$
for all $t$.
Thus, the identity of bounded operators
$W(t)=G\Gamma^{4it}G$ holds for all $t$. By assumption we have
$\wlim_{t\to0}W(t)=E$, thus 
$W(s)W^\ast(s)=G\Gamma^{4is}\Gamma^{-4is}G=G$ implies $G=E$.
\end{proof}
\begin{Llemm}{auto}
  The action 
$\tau_t^E\colon\OA_E\ni A_E\longmapsto\tau^E_t(A_E)=\Gamma^{4it}A_E\Gamma^{-4it}$
is a strongly continuous group of automorphisms of $\OA_E$.
\end{Llemm}
\begin{proof}
  For $A_E=EAE\in\OA_E$ we have $E\Delta^{\ii t/n}E A_E E\Delta^{-\ii t/n}E=
E\sigma_{t/n}(A_E)E$, 
where $\sigma$ is the modular group of $(\OA,\Omega)$, and this 
shows $F_n(t)A_EF_n(-t)\in\OA_E$ for all $n$. 
Since $\OA_E$ is weakly closed
and $F_n(t)A_EF_n(-t)$ converges strongly, and therefore also weakly, by 
assumption~i) of \refT{modular}, it converges to an element of $\OA_E$.
Since $\NORM{F_n(t)A_EF_n(-t)}\leq\NORM{A_E}$ for all $n$,
the limit mapping is continuous
on $\OA_E$. By \refL{Boundary}, it equals $\tau^E_t$, as defined above,
for all $t$. Since $\Gamma^{4it}$ is a strongly continuous group of 
unitary operators on $E\HS$, the assertion follows.
\end{proof}
\begin{proof}[Proof of \refT{modular}]
We note first, that $W(-t)=W(t)^\ast$ can be seen by direct methods
as in~\cite{MS77}. Secondly, since $\tau^E$ is an automorphism group
of $\OA_E$, it follows by definition of $\HS_E$, that the $W(t)$ leave
that subspace invariant and thus form a unitary group on it.
The stated analyticity properties of $W$ are contained in the conclusions
of Lemmas~\ref{lemm:HoloLim} and~\ref{lemm:Boundary}.
\end{proof}
\subsection{Preliminary Conclusions and Remarks}
\label{sec:conc1}
Let us pause for a moment to put the results compiled so far into perspective.
Reconsidering the assumptions and conclusions of Theorems~\ref{theo:unitary}
and~\ref{theo:modular}, 
one might wonder what has actually been proved. In fact, the only 
additional information gained is the group property of the boundary values $T$,
respectively, $W$. That is, the
theorems ensure the existence of the \textit{Zeno dynamics} if it is
already known that the \textit{Zeno effect} occurs, and persists in the Zeno limit. 
Both results therefore have a quite different status from
\refT{analytic-semigroups}.
Hence the results of Sections~\ref{sec:unitary}
and~\ref{sec:modular} have to be complemented by convergence
conditions that can be efficiently tested in models. This will be the
subject of the remainder of the present paper. Yet, we use the occasion to 
present some general thoughts in that direction.

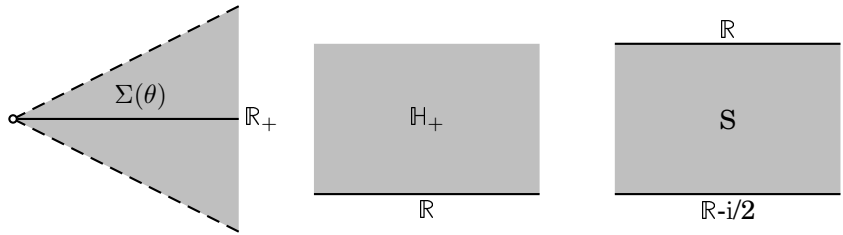
\begin{figure}[htbp]
  \centering
\newgray{mid1}{0.75}
\newgray{mid2}{0.45}
  \psset{xunit=1mm}%
\psset{yunit=1mm}%
\begin{pspicture}(0,0)(110,30)
\pspolygon[linestyle=none,fillstyle=solid,fillcolor=mid1](0,15)(30,30)(30,0)(0,15)
\psline[linestyle=dashed,linecolor=black](0,15)(30,30)
\psline[linestyle=dashed,linecolor=black](0,15)(30,0)
\psline[linestyle=solid,linecolor=black]{o-}(0,15)(30,15)
\pspolygon[linestyle=none,fillstyle=solid,fillcolor=mid1](40,5)(40,25)(70,25)(70,5)
\psline[linestyle=solid,linecolor=black](40,5)(70,5)
\pspolygon[linestyle=none,fillstyle=solid,fillcolor=mid1](80,5)(80,25)(110,25)(110,5)
\psline[linestyle=solid,linecolor=black](80,5)(110,5)
\psline[linestyle=solid,linecolor=black](80,25)(110,25)
\rput{0}(95,3){\RR-\ii/2}
\rput{0}(95,27){\RR}
\rput{0}(55,3){\RR}
\rput{0}(95,15){S}
\rput{0}(55,15){$\mathbb{H}_+$}
\rput{0}(33,15){$\RR_+$}
\rput{0}(17,18){$\Sigma(\theta)$}
\end{pspicture}
  \caption{Analyticity domains relevant for Zeno dynamics}
  \label{fig:analytic}
\end{figure}

We have seen in Section~\ref{sec:semigroup} 
that the occurrence of the Zeno effect does not hinge on unitary
evolutions, as has been noted by various authors in different
contexts~\cite{PL98,WHI00}. 
Rather, for the existence of Zeno dynamics\ie the infinitely frequent
measurement limit, analyticity properties of the original evolution seem to be of
the essence. Three fundamentally different cases can be distinguished, and the
pertinent analyticity domains are sketched in Figure~\ref{fig:analytic}.
In the simplest case of open sectors, shown on the left hand side,  
convergence of the Zeno dynamics is virtually unconditional. 
This was used in Section~\ref{sec:unitary} to conclude that
the Zeno dynamics exists in the upper half-plane $\mathbb{H}_+$, to simplify
the proof of Misra's and Sudarshan's Theorem. 
In the third case of von Neumann Algebras, the sectorial result could
not be used at all since the domain of analyticity does not contain
an open sector, and we were bound to use the original
`bootstrap' strategy for the proof. 
But what renders these latter two cases more difficult is obviously the 
necessity to extend the limiting dynamics from the interior of the 
analyticity domain to its boundary.
The general technique that is always applied here is the usage of reproducing
kernels like the Cauchy--Hilbert or Poisson kernel to render a function
as the boundary value of some complex integral transform of itself~\cite{b:BRE65}.
The difference described above between the first and the other two cases is also the reason 
why the latter do not yield explicit conditions for the existence of Zeno 
dynamics, but have to assume convergence \textit{a priori}. 

In the theory of generalised functions,
many conditions are known as to when and in which sense a boundary value of an 
analytic function can be taken, and constitutes a continuous function
(smooth function, tempered distribution, hyperfunction). These conditions
hinge on the asymptotic growth of the analytic function as the boundary 
of the domain is approached --- for continuity, it simply has to stay bounded.
In turn, such growth conditions can be derived if the analytic function
is considered as the Fourier--Laplace transform of some function on the real 
axis, and the properties of the boundary values then depend essentially on
the growth or decay of the latter at infinity. Characterisations of regularity of 
functions by  the growth of their (inverse) Fourier--Laplace transform are
known as Payley--Wiener Theorems, see~\cite{b:RS72,b:RUD74,KAW70,AUS02C}, 
the simplest instance of which
is the well known Riemann--Lebesgue Lemma, stating that the Fourier 
transforms of continuous functions vanishing at real infinity 
form exactly the class of integrable functions~\cite[Theorem~7.5]{b:RUD74}. 

Now, a unitary quantum evolution $U(t)$
generated by a Hamiltonian $H$ is the Fourier--Laplace 
transform of the Hamiltonian's spectral density $P(\lambda)$,
\[
U(z)=\ee^{\ii z H}=\int_\RR \ee^{\ii z \lambda} \dd P(\lambda).
\]
The growth conditions on $P$ were what allowed us to extend
$U$ to the upper half-plane in the quantum mechanical case
--- $P$ supported in $\RR_+\cup\{0\}$ since $H$ is semibounded
--- respectively, to a strip in the von Neumann  
case, where $P$ is exponentially damped in one direction. 
One should note that the semiboundedness
of the Hamiltonian in the quantum mechanical case is \textit{not}
a necessary precondition for the Zeno effect, as was previously
thought~\cite{MS77}. This is clearly shown by Theorem~\ref{theo:modular}.

In conclusion, it would hence seem tempting to formulate
growth conditions on some spectral density which would ensure the
existence of Zeno dynamics, all the more since the Payley--Wiener Theorems
usually provide \textit{necessary and sufficient} conditions, due to
the continuous invertibility of the Fourier--Laplace transformation. Yet,
the spectral density one should consider here would be that of the 
generator of the desired Zeno dynamics itself, and that object
is generically unknown \textit{a priori}. In essence, the desired 
conditions would depend on detailed information about the energy
distribution, with respect to the original Hamiltonian, of the states
in the Zeno subspace\ie one of the (possibly many), invariant subspaces for the
Zeno dynamics. Results in this promising direction are as of yet scarce,
some remarks can be found in Section~2 of~\cite{NNP96}.
We will present an example in Section~\ref{sec:EDF}, where projections
of rank one are considered, and conditions on the energy density distribution
of the decaying initial state are given which are equivalent to the existence of
the Zeno limit. 

We conclude this section with some remarks of a more technical nature.
\begin{rem*}[\textbf{On Semigroups with Bounded Generators}]
  We omitted from consideration the simple case of $C_0$-semigroups
with bounded generators, for which a convergence result was noted
in~\cite{MAS02}.
\begin{theo}[{\cite[Theorem~2.1]{MAS02}}]
  Let $A$ be a bounded operator on a Banach space $\mathcal{B}$,
generating a $C_0$-semigroup $(\ee^{tA})_{t\geq0}$, and $E$ a bounded
projection on $\mathcal{B}$.
Then
\[
\lim_{n\to\infty} \Bigl[ \ee^{\frac{t}{n}A} E\Bigr]^n\phi =
\lim_{n\to\infty} \Bigl[ E\ee^{\frac{t}{n}A} \Bigr]^n \phi =
\lim_{n\to\infty} \Bigl[ E\ee^{\frac{t}{n}A}E \Bigr]^n\phi  =
\ee^{tEAE}\phi, 
\]
for all $\phi\in\mathcal{B}$ and uniformly in $t\in[0,T]$ for all $T\geq0$. 
\end{theo}
This is another case in which the Zeno paradox emerges unconditionally
(in the sense of the following remark),
but here it is also directly possible to identify the generator of
the Zeno dynamics as $EAE$, a result for which additional conditions
are needed in the general case of unbounded generators, see 
Sections~\ref{sec:gen-equi} and~\ref{sec:EI}.
The direct proof in~\cite{MAS02} is based on Chernoff's product 
formula~\cite{b:CHE74}.
It uses the decomposition $\phi=E\phi+(\bbone-E)\phi$
iteratively in its estimations, in a manner that is similar to the method
that was used independently in~\cite{AUS03}, 
and will be used in \refP{AZC}, 
to obtain a sufficient condition for the 
existence of the Zeno limit. 
\end{rem*}
\begin{rem*}[\textbf{On Two Complete Characterisations}]
It recently turned out that the two cases of sectorial semigroups, and such with
bounded generators are in fact particularly benign~\cite{MAT03}. They are in fact,
in the Hilbert space case,
completely characterised by the 
unconditional
emergence of the Zeno paradox.
Here unconditional means in any possible case\ie for every projection.
\begin{theo}[{\cite[Theorem~3]{MAT03}}]
  Let $A$ be the generator of the  $C_0$-semigroup $(\ee^{tA})_{t\geq0}$
on the Hilbert space \HS. Then $A$ is bounded ($-A$ is associated with a 
densely-defined, closed, sectorial form) if and only if, for all $\phi\in\HS$, $t>0$,
\[
\lim_{n\to\infty} \Bigl[ \ee^{\frac{t}{n}A} E\Bigr]^n\phi
\]
exists for all bounded projections (orthogonal
projections) $E$.
\end{theo}
What this complete characterisation also says is that there
will be counterexamples in the
quantum mechanical, and all the more in the von Neumann case where
the generic generator is only semibounded or even unbounded, and 
the associated semigroup
cannot
be extended to a sector around the real axis. That is, there will be
some orthogonal projections for which the Zeno paradox does not emerge.
Very recently, 
Matolcsi also reported on the analogous results
in the Banach space case~\cite{MAT03A}.
\end{rem*}
\begin{rem*}[\textbf{On CPT symmetry}]
  A slight weakening of the assumptions of \refT{unitary} is possible if
the system under consideration possesses a time reversal or, more 
specially, a CPT symmetry\ie an antiunitary
operator $\theta$ such that
\begin{align*}
  & \theta E\theta^{-1}=E,\\
  & \theta U(t) \theta^{-1} = U(-t)\quad\text{for all } t.
\end{align*}
Then it suffices to assume existence of the limit $T(t)$ only for 
$t>0$, since
\begin{align*}
  T(-t) &= \slim_{n\to\infty} \bigl[ EU(-t/n)E\bigr]^{n}\\
&=  \slim_{n\to\infty} \theta\bigl[ EU(-t/n)E\bigr]^{n}\theta^{-1}
= \theta T(t) \theta^{-1}.
\end{align*}
Although this simplification can usually be applied in the case of quantum 
systems --- where CPT symmetry is generic --- it is not all-too helpful
in concrete models, since the conditions used there imply 
convergence on the whole axis anyway, see~\refS{Conditions}.
\end{rem*}
\begin{rem*}[\textbf{On the Zeno Subspace}]
  In Sections~\ref{sec:semigroup} and~~\ref{sec:unitary} we found that
the Zeno dynamics is confined to the subspace $E\HS$. 
Yet, this is only the \textbf{\textit{maximal} Zeno subspace} that is
invariant under the Zeno dynamics, if it exists at all. And in fact
we saw in \refT{modular} that there the Zeno dynamics can be confined
to the generally smaller subspace
\[
\HS_E=\overline{\OA_E\Omega}=\overline{E \OA E\Omega} \nsubseteq 
\overline{E \OA \Omega}=E\HS.
\]
The main reason that made this identification possible was that
$E$ was assumed to be an element of the von Neumann algebra \OA in question\ie
using the terminology of algebraic quantum theory, 
that it was an \textit{observable}. This seems to be a natural choice
in this context for physical reasons, as can be seen in the models
considered in~\cite{AUS03}, and it is also most closely related to
the colloquial description of the Zeno effect as `evolution interrupted
by frequent measurement'. But in the general case of arbitrary projections
on a separable Hilbert space, larger Zeno subspaces can appear, 
and have been characterised in~\cite{EI03}. We will come back
to that issue in Section~\ref{sec:EI}.
\end{rem*}
\begin{rem*}[\textbf{On Separability}]
The conditions of weak continuity in the two Theorems~\ref{theo:unitary}
and~\ref{theo:modular}, can be relaxed if the Hilbert space considered
is separable.
To make the argument clear, consider
for instance \refT{unitary},  where the following reasoning
can be used to conclude that $W(t+\ii\eta)$ weakly approximates $T(t)$, if
\HS is separable.
We saw that 
\[
\lim_{\eta\to0}\IPROD{\phi}{W(t+\ii\eta)\psi}=\IPROD{\phi}{T(t)\psi}
\]
$t$-almost everywhere
\ie
outside a set $\mathcal{N}_{\phi,\psi}$ of Lebesgue measure zero.
For a countable, dense set $\mathcal{D}\subset\HS$ (and here is the only
place where separability of \HS is used), set 
$\mathcal{N}=\cup_{\phi,\psi\in\mathcal{D}}\mathcal{N}_{\phi,\psi}$, which is
a null set as the countable union of null sets. Now, approximate two arbitrary
vectors $\phi$, $\psi\in\HS$ by sequences $\{\phi_n\}_{n\in\NN}$,
$\{\psi_n\}_{n\in\NN}\subset\mathcal{D}$. With
$A(s,\eta)=W(s+\ii\eta)-T(s)$ we have
\[
\IPROD{\phi}{A(s,\eta)\psi}=
\IPROD{\phi-\phi_n}{A(s,\eta)\psi}+
\IPROD{\phi_n}{A(s,\eta)(\psi-\psi_n)}+
\IPROD{\phi_n}{A(s,\eta)\psi_n}
\]
For $s$ outside $\mathcal{N}$, 
the third term tends to zero as $\eta\to0_+$, 
since $\phi_n$, $\psi_n\in\mathcal{D}$, while the first
and second converge to zero as $n\to\infty$.
Together, this shows 
$\wlim_{\eta\to 0_+} W(t+\ii\eta) = T(t)$, 
$t$-almost everywhere, from which point one can proceed as in~\cite{MS77}.
Essentially the same argument applies to \refT{modular} when the
GNS--Hilbert space is separable, and this is in fact the prevalent case 
for models in quantum statistical mechanics, in which the states inducing
the representations are, for instance,
constructed as thermodynamic limits of locally normal states, cf.~\cite{b:BR79/81}.
In particular in the von Neumann
case  we assumed this state to be normal, which implies 
separability, and thus the assumptions of \refT{modular}
are actually a bit too strong.
\end{rem*}
%
\section{Operator Theoretical Conditions}
\label{sec:Conditions}
\subsection{The Asymptotic Zeno Condition}
\label{sec:AZC}
The first condition for the existence of the Zeno limit we
want to present was found in~\cite{AUS03}, and applied to
quantum statistical mechanics. It is closely related
to the quadratic short-time behaviour of quantum evolutions,
which is commonly associated with the Zeno effect~\cite{AA90,NNP96}, 
see Section~\ref{sec:geometry}.
We present it in a more neutral
form which makes clear that its scope is somewhat broader.

\begin{defi}
  Let $E$ be a projection on a Hilbert space \HS and $(U(t))_{t\in\RR}$
a unitary group on \HS.
 We say that $(U,E)$ satisfies the
\textbf{uniform (strong) asymptotic Zeno condition}, in short,
\textbf{uAZC (sAZC)} if the asymptotic relation
\[
E^\perp U(\tau)E=O(\tau) \quad \text{uniformly (strongly) as $\tau\to0$},
\]  
holds. This relation
means that there shall exist $\tau_0>0$ and $C\geq0$ such that for all $\tau$ with 
$\ABS{\tau}<\tau_\psi$ holds the estimate
$\NORM{E^\perp U(\tau)E}\leq C^{1/2}\ABS{\tau}$ (respectively, 
for any $\psi\in\HS$ exist $\tau_0>0$ 
and $C_\psi\geq0$ such that for all $\tau$ with 
$\ABS{\tau}<\tau_0$ holds  
$\NORM{E^\perp U(\tau)E\psi}\leq C_\psi^{1/2}\ABS{\tau}$).
\end{defi}
The (uAZC) is in fact equivalent to saying that the function $E^\perp U(t) E$ is
uniformly Lipschitz continuous at the point $t=0$. Not surprisingly,
Lipschitz continuity is well known as a salient condition for the 
existence of solutions to (nonlinear) evolution equations.
For us, it ensures existence of the evolution in the Zeno limit.
\begin{Lprop}{AZC}
 If $(U,E)$ satisfies uAZC (sAZC) then
\[
F_n(t)\DEF \bigl[E U(t/n) E\bigr]^n, \quad
\text{for }t\in\RR,\ n\in\NN,
\]
converge uniformly (strongly) to operators $W(t)$.
Furthermore $W(t)$ is 
uniformly (strongly) continuous in $t$ and  the uniform 
(strong) limit as $t\to0$ of $W(t)$ is $E$.
\end{Lprop}
\begin{proof}
  We present the proof of the 
uniform case of which the strong one is a 
straightforward generalisation.
We want to see whether the $F_n(t)$ form a Cauchy sequence in $n$ for given $t$.
For that, we have to estimate the quantities
\[
  \NORM{\bigl(F_n(t)-F_m(t)\bigr)}\leq
  \NORM{\bigl(F_n(t)-F_{nm}(t)\bigr)} + 
  \NORM{\bigl(F_m(t)-F_{nm}(t)\bigr)}.
\]
A double telescopic estimation yields
\begin{multline*}
  \NORM{\bigl(F_n(t)-F_{nm}(t)\bigr)} \leq\\
  \sum_{k=1}^{n}\sum_{l=1}^{m-1}
  \left\|
  \bigl[EU(t/n)E\bigr]^{n-k} 
  \left( EU(t(m-l)/(nm))E \bigl[EU(t/(nm))E\bigr]^{l} -\right.\right.\\
   \left.\left. EU(t(m-l+1)/(nm))E \bigl[EU(t/(nm))E\bigr]^{l-1} \right)
   \bigl[EU(t/(nm))E\bigr]^{m(k-1)}\right\| .
\end{multline*}
Now, since with $E^{\perp}\DEF\bbone-E$ we have
\[
EU(t(m-l+1)/(nm))E = E U(t (m-l)/nm)(E+E^\perp )  U(t/(nm)) E, 
\]
we find that the $(k,l)$th term in the sum is equal to
\begin{multline*}
  \left\|
  \bigl[EU(t/n)E\bigr]^{n-k} 
  \cdot EU(t(m-l)/(nm)) E^\perp \cdot \right.\\
  \left. \cdot E^\perp U(t/(nm)) E \cdot \bigl[EU(t/(nm))E\bigr]^{l-1}
   \bigl[EU(t/(nm))E\bigr]^{m(k-1)}\right\| .
\end{multline*}
Multiplying out and using repeatedly $\NORM{AB}\leq\NORM{A}\NORM{B}$, we estimate this expression
from above by
\begin{multline*}
  \NORM{\bigl[EU(t/n)E\bigr]}^{n-k} 
  \cdot \NORM{EU(t(m-l)/(nm)) E^\perp}\\
  \cdot \NORM{E^\perp U(t/(nm)) E }\cdot \NORM{\bigl[EU(t/(nm))E\bigr]}^{l-1}
   \NORM{\bigl[EU(t/(nm))E\bigr]}^{m(k-1)} .
\end{multline*}
Observing that all terms containing only the projection $E$ have operator norm $\leq1$
and can thus be omitted in the estimation of $\NORM{\bigl(F_n(t)-F_{nm}(t)\bigr)}$,
we arrive at
\[
  \NORM{\bigl(F_n(t)-F_{nm}(t)\bigr)} \leq  
   \sum_{k=1}^{n}\sum_{l=1}^{m-1}
  \NORM{E U(t(m-l)/(nm)) E^\perp}\NORM{ E^\perp U(t/(nm)) E }.
\]
Now, for $n>n_0\geq1/\tau_0$, and $m\geq2$,  the uAZC implies
\begin{align*}
  \NORM{\bigl(F_n(t)-F_{nm}(t)\bigr)} &\leq
  C
  t^2\sum_{k=1}^{n}\sum_{l=1}^{m-1}
  \frac{m-l}{n^2m^2}\\& =
  C
t^2\sum_{k=1}^{n} \frac{(m-1)m}{2n^2m^2}\\ & =
  \frac{C
t^2}{2} \frac{(m-1)m}{n m^2}\leq 
  \frac{C
t^2}{2n}.
\end{align*}
An analogous estimate holds for
$\NORM{\bigl(F_m(t)-F_{nm}(t)\bigr)
}$, which yields
for $m-2\geq n>n_0\geq1/\tau_0 $ the overall result
\begin{equation}
  \label{eq:AZCest}
 \NORM{\bigl(F_n(t)-F_m(t)\bigr)}\leq \frac{Ct^2}{n}.  
\end{equation}
The first statement of this 
proposition is now clear since the estimate above shows
that the $F_n(t)$ form Cauchy sequences which are therefore
\textit{a fortiori} convergent. The other statements follow
from 
$F_n(0)=E$ for all $n$, and the fact that the convergence 
of $F_n(t)$ is uniform for $t$
on compact subsets of \RR. This follows in turn from the
$t$-dependence of the final estimate.
\end{proof}
Combining this result with \refT{unitary}, we immediately
obtain an efficient condition for Zeno effect and dynamics in the
quantum mechanical case.
\begin{coro}
Let $U(t)=\ee^{\ii tH}$ be a unitary group
with nonnegative, self-adjoint generator $H$.
If $(U,E)$ satisfies uAZC or sAZC then
$W(t)=\smash[b]{\slim\limits_{n\to\infty}}F_n(t)$ 
is a degenerate group of unitaries on $E\HS$.
\end{coro}
The AZC can be related to one well known, basic condition
for the Zeno paradox in quantum mechanics, namely finiteness of the first
moment of the Hamiltonian in the initial state. Consider
the rank one projection $E_\psi$ onto a vector in the domain
of $H$, and denote by 
$\OA(t)\DEF\IPROD{\psi}{\ee^{\ii tH}\psi}$
the \textbf{survival amplitude} in this state. 
Then holds the elementary asymptotic expansion
\[
\OA(\tau)\sim 1+\ii\tau\IPROD{\psi}{H\psi}-\frac{\tau^2}{2}\IPROD{\psi}{H^2\psi}
\quad (\tau\to 0),
\]
with
\[
\IPROD{\psi}{H\psi}<\infty, \quad \IPROD{\psi}{H^2\psi}=\NORM{H\psi}^2<\infty.
\]
This is, as will be seen in Section~\ref{sec:geometry},  
a rather direct expression of the quadratic short time behaviour of
quantum dynamics, 
which is commonly identified as the main cause for Zeno effect
and paradox~\cite{NNP96,WHI00,FAC02}. 
Obviously, it implies uAZC and therefore ensures convergence to the Zeno limit.

The main strength of the AZCs lies in the fact that they can
be
efficiently 
tested in concrete models. In particular they enable the use of 
perturbation theory to obtain conditions for Zeno dynamics. This has been 
used in~\cite{AUS03} to treat models of quantum statistical mechanics
accordingly, and we will present two more generic of the pertinent
results below in \refS{AZC-app}.
The AZCs are quite weak and thus 
indicate how generic a quantum phenomenon the Zeno effect indeed is.
For example it is always satisfied if the generator $H$ of the group $U$
is bounded, or, more generally, if $E$ projects onto
a closed subspace of entire analytic elements for $H$\eg
if $E$ is contained in a bounded spectral projection of $H$. 
In those cases a power series expansion of $U(t)=\ee^{\ii t H}$
implies uAZC. However, if neither is the case, then uAZC will generally 
fail to hold in that its defining estimate is not 
uniform in $\psi\in\HS$, respectively, sAZC will not hold for
all $\psi\in\HS$ (not even on a dense subset).
It is also noteworthy that in showing the convergence of $F_n$ to $W$,
we have not used the unitarity of $U$. Thus an analogue of the Zeno
effect is also possible for non-unitary (%
non-Hamiltonian,
non-Schrödinger) evolutions, as already noted above. 
On the other hand, the group property of $U$ was essential for 
obtaining the quadratic term that forced
the convergence of the sequence.
Asymptotic bounds on $E^\perp U(t) E$ have already been considered 
by other authors~\cite{MS77A,NIS88, EX89}, in the context of short-time
regeneration of an undecayed state. In particular in~\cite{EX89}, the
deviation of the `reduced evolution' $EU(t)E$ from being a semigroup has 
been expressed by such (polynomial) bounds. We will obtain a similar
yet somewhat coarser result in \refS{AZC-app}, and present a much more
advanced one in \refS{EI}.
\subsection{Application of the AZC to Quantum Statistical Mechanics}
\label{sec:AZC-app}
One of the most successful mathematical theories for physical phenomena
is the algebraic formulation of quantum statistical mechanics and
quantum field theory~\cite{b:BR79/81,b:HAA92}. Its basic tenet is the viewpoint that all
relevant information about a system resides in its \textbf{observable algebra}
\OA, a topological algebra which captures all \textit{finite} measurements
that can be performed on the system, where finiteness is to be understood with
respect to time, space, and energy resources. In this context it became
clear that \textit{weakly closed}\ie $W^\ast$ or von Neumann algebras
are the natural objects to consider, and this revealed another deep 
connection between `pure' mathematics and physics. In particular the theoretical
development of quantum statistical mechanics was spurred by the close
relation between the modular dynamics of von Neumann algebras and the
notion of thermal equilibrium states of $C^\ast$, respectively, 
$W^\ast$-dynamical systems, incorporated in the \textbf{KMS condition}.
This connection, given by Takesaki's theorem~\cite[Theorem~5.3.10]{b:BR79/81},
is the fundament for the application of the mathematical result \refT{modular}
to a general $W^\ast$-dynamical system 
$(\OA,\tau)$ with faithful, normal KMS state $\omega$ at
\textbf{inverse temperature} $\beta$ (termed $(\tau,\beta)$--KMS state).
This was laid out in~\cite{AUS03}, and we report on the central results
in the present section.

To fix the context, we denote by $\Omega$ the vector representative
of $\omega$ in the associated representation 
$\pi_\omega$ on the (separable) GNS--Hilbert space $\HS$.  
The automorphism group $\tau$ is assumed to be implemented covariantly\ie
by a strongly continuous group of unitary operators $U(t)$
on $\HS$. The representation
$\pi_\omega$ will be omitted from the notation, 
when no confusion is possible. We are now ready to combine
\refT{modular} and the AZC to obtain an effective condition for the 
emergence of Zeno dynamics in quantum statistical mechanics.
\begin{Ltheo}{ZDexist}
Under the conditions described above, let $\beta>0$, assume $\OA$ to be unital, 
let $E\in\OA$ be a projection, and set $E^\perp\DEF\bbone-E$. Assume that
the \textbf{asymptotic Zeno condition} holds:
For $A\in\OA$, the estimate
\[
\NORM{E^\perp U(\zeta) E A\Omega} \leq C\cdot\NORM{A\Omega}\cdot \ABS{\zeta}
\]
is valid for $\zeta\in\CC$ with $\ABS{\zeta}<r_0$ for some fixed 
$r_0>0$ and $\IM\zeta\geq 0$. 
In short: $(U,E)$ satisfies AZC for $\OA$.
Then the strong operator limits
\[
W(t)\DEF\slim_{n\to\infty}
\bigl[E U(t/n) E\bigr]^n
\]
exist, and form a strongly continuous group of unitary operators
on the \textbf{Zeno subspace} $\HS_E\DEF\overline{\OA_E\Omega}\subset E\HS$,
where $\OA_E\DEF E\OA E$. The group $W(t)$
induces an automorphism group $\tau^E$ of  
$\OA_E$, 
such that $(\OA_E,\tau^E)$ is a 
$W^\ast$-dynamical system. The vectors 
$W(z)A_E\Omega$,  $A_E\in\OA_E$, extend analytically 
to the strip $0<\IM{z}<\beta/2$ and 
are continuous on its boundary.
\end{Ltheo}
The slight modification of the AZC is needed here, because to satisfy the
conditions of \refT{modular}, we need convergence on \textit{both}
boundaries of the strip $\{0\leq\IM z\leq\beta/2\}$ in the complex plane,
which would in general not hold if we assumed AZC only 
on
the real axis.
As was shown in~\cite{AUS03}, this version of AZC is satisfied in many model
cases, of which we will present a generic one further down in this section.
\begin{proof}
We show that the assumptions 
of \refT{modular} are satisfied, 
from which we obtain the stated conclusions.
First, for real $\tau$, the AZC implies $E^\perp U(\tau) E=O(\tau)$ 
uniformly since the operators in question are bounded, $\OA\Omega$ is dense in $\HS$,
and AZC holds uniformly in $A$ on a fixed real neighbourhood of $0$. 
Therefore \refP{AZC} yields the existence
of $W(t)$, $t\in\RR$, its weak continuity in $t$ and the initial condition
$\wlim_{t\to0}W(t)=E$. These facts comprise condition~i)
of \refT{modular}.
For the second condition of the cited theorem, we need only to show
that $W(t+\ii\beta/2)$ exist as strong operator limits on the common, dense
domain $\OA\Omega$.
For this notice that the calculations yielding
\refE{AZCest}
are applicable to $(\smash{F_n(t+\ii\beta/2)-F_m(t+\ii\beta/2)})A\Omega$, 
leading to the estimate
\[
 \NORM{\bigl(F_n(t+\ii\beta/2)-F_m(t+\ii\beta/2)\bigr)
A\Omega
}\leq
 \frac{C
 \NORM{A}
\ABS{t+\ii\beta/2}^2}{n}
\]
for $A\in\OA\Omega$, and $m-2\geq n> n_0\geq 1/r_0$.
Thus, also condition~ii) of \refT{modular} is
satisfied and the stated conclusions follow from it.
\end{proof}
It should be noted that we 
restrict our discussion completely to a concrete 
realisation
of a  $W^\ast$-dynamical system given by the GNS
representation $\pi_\omega$ of a fixed, \textit{a priori} chosen 
KMS state $\omega$. That is we consider the 
von Neumann algebra 
$\pi_\omega(\OA)$ on the GNS Hilbert space \HS and assume the 
dynamical automorphism group to be $\pi_\omega$-covariant\ie to 
be 
realised
by a strongly continuous, unitary group of operators. 
This notably simplifies our treatment, but also restricts it
to a single superselection sector of the theory. Nevertheless,
the results in Section~\ref{sec:gen-equi}
below are essentially independent of the chosen representation.

\refT{ZDexist} leads  directly to what may be seen as \textit{the}
proper manifestation of the Zeno paradox in quantum statistical
mechanics, and what is the closest counterpart to
the Zeno paradox in quantum mechanics\ie
the prevention of a decay process~\cite{HNNPR98,FGMPS00}.
In the present context, the Zeno effect can prevent the
\textit{return to an equilibrium state}, as we will show now.

As said, The power of the AZC lies to a great extent in the fact that it
yields perturbative conditions for the occurrence of the Zeno effect.
For it is known that a perturbed semigroup $U^P_t$, resulting from adding
a bounded perturbation $P$ to a $C_0$-semigroup $U_t$, is close to $U_t$
for small times in the sense that
$\NORM{U_t-\smash{U^P_t}}=O(t)$, as $t\to0$, 
see~\cite[Theorem~3.1.33]{b:BR79/81}. Now if
$E$ projects onto a subspace which is invariant under $U_t$, then
this asymptotic behaviour implies that the Zeno dynamics of the 
pair $(U^P_t,E)$ exists. We exemplify this basic mechanism in
the following.

It is well known~\cite{ROB73} that a quantum system will under
general conditions\eg if $(\OA,\tau)$ is asymptotically Abelian, 
return to equilibrium for large times. 
This means that if the system is prepared in an equilibrium state $\omega^P$
for the perturbed evolution $\tau^P$, 
where $P=P^\ast\in\OA_\tau$ is a bounded perturbation, which is
in the set of entire analytic elements $\OA_\tau$ 
for $\tau$ (termed local perturbation), 
and thereafter evolves under the unperturbed 
dynamics $\tau$, one recovers a $\tau$-equilibrium state $\omega_±$ 
for $t\to±\infty$, in the weak* topology. 
Assume that the perturbed and unperturbed dynamics
are implemented by unitaries $U^P$ and $U$, respectively, which
is always possible if either $\tau$ or $\tau^P$ 
is covariant in the chosen representation~\cite[Theorem~1]{ROB73}.
Then, the unperturbed dynamics can be written
in terms of the perturbed one by the perturbation
expansion~\cite[Theorem~3.1.33 and Proposition~5.4.1]{b:BR79/81}
\[
      U(t)=U^P(t)+ 
      \sum\limits_{n\geq1}\int\limits_0^t\dd t_1\dotsi\int\limits_0^{t_{n-1}}\dd t_n
      U^P(t_1) P U^P(t_2-t_1)P \cdots P U^P(t-t_n), 
\]
where the $n$-th term in the sum is bounded by $\NORM{P}^nt^n/n!$.
Let the system be prepared
in any $\tau^P$-invariant state $\varphi^P$. 
In the representation $\pi^P$
induced by the chosen $\tau^P$-KMS state $\omega^P$
the corresponding vector states are denoted by
$\Phi^P$ and $\Omega^P$ respectively. Let $E$ be the projection onto
the space spanned by the vector $\Phi^P$ and assume $E\in\OA$.  Then the above expansion
readily yields $E^\perp U(t) E=O(t)$ uniformly,
since the $\tau^P$-invariance of $\varphi^P$ 
implies $U^P(t)\Phi^P=\Phi^P$. In application to vectors in
$\OA\Omega$ this estimate extends to a fixed, small neighbourhood
of $0$ in the upper half-plane and is uniform in those vectors.
Thus the AZC holds, the Zeno dynamics converges, and
the system remains in the state $\varphi^P$. The same reasoning
is applicable if $E$ projects onto a $\tau^P$-invariant subspace.
Let us resume what we have proved.
\begin{Lcoro}{non-return}
  Let $(\tau,\OA)$ be as above. 
Let $P\in\OA$ be a local perturbation,
and denote by $\tau^P$
perturbed dynamics as constructed 
in~\cite[Proposition~5.4.1 and Corollary~5.4.2]{b:BR79/81}.
Let $E\in\OA$ be a $\tau^P$-invariant projection\ie $\tau^P(E)=E$. 
Then the $(\tau,E)$-Zeno dynamics
$\tau^E$ is an automorphism group of 
$\OA_E$, and $\HS_E$ is $\tau^{E}$-in­va­ri­ant.
\end{Lcoro}
\subsection{Zeno Generator and Zeno Equilibria}
\label{sec:gen-equi}
Let us remain for another while in the context of quantum
statistical mechanics to emphasise how favourable its
mathematical framework is for the study of the Zeno effect.
In particular we want to see that the AZC can be used to
identify the generator of the Zeno dynamics explicitly.
Let $H$ be the generator of $U(t)=\ee^{\ii tH}$. 
The unitary group $U_E(t)\DEF\ee^{\ii t EHE}$ is 
called the \textbf{reduced dynamics} associated with $(U,E)$, whenever it
is defined on the Zeno subspace $\HS_E$.
To be able to compare the reduced with the Zeno dynamics, we need
a technical condition, which has been shown in~\cite{AUS03} to be
satisfied in many models where also the AZC holds:
We call $(U,E)$ \textbf{regular} if $\OA_E$ contains a dense
set of elements which are analytic for $\tau$ in an arbitrary
neighbourhood of zero. The condition of regularity will be
required to have enough analytic vectors in $\HS_E$ at hand 
for the proof below to work. It excludes pathological cases\eg 
when $E$ projects onto a subspace of states with properly infinite energy.
\begin{Lprop}{reduced}
  Let $(U,E)$ be regular and satisfy AZC for \OA. 
  Then $U_E(t)$ equals $W(t)$, when restricted to $\HS_E$.
\end{Lprop}
  Throughout the proof below let $\psi_E\in \OA_{E,\tau}\Omega\subset\HS_E$, where
$\OA_{E,\tau}$ is a dense set of  elements in $\OA_E$, which are analytic for $\tau$.
Record that, by the discussion following~\cite[Definition~3.1.17]{b:BR79/81}, 
the $\tau$-analyticity
of $\psi_E$ is equivalent to analyticity with respect 
to $U$ and this is in turn equivalent to the convergence of power series
of analytic functions in $\sigma H$ applied to $\psi_E$, 
for $\sigma\in\CC$ small enough,
as given in the cited definition.
\begin{proof}[{Proof of~\refP{reduced}}]
We first derive a useful asymptotic estimate:
Setting $\psi_E(\sigma)\DEF U_E(\sigma)\psi_E$ holds 
\begin{align*}
  \NORM{\bigl(U_E(\tau) - EU(\tau)E\bigr)\psi_E(\sigma)} &= 
  \NORM{ \left\{ \sum_{k=0}^\infty \frac{(\ii\tau)^k (EHE)^k}{k!}   - 
        E \sum_{l=0}^\infty \frac{(\ii\tau)^l H^l}{l!} E \right\} \psi_E(\sigma)} \\
  &= \NORM{\sum_{k=2}^\infty\frac{(\ii \tau)^k}{k!} \bigl[ (EHE)^k  - EH^kE \bigr]\psi_E(\sigma)},
\intertext{using $E\psi_E(\sigma)=\psi_E(\sigma)$, 
which is clear since $U_E$ commutes with $E$. 
By using $\NORM{E}=1$, this can be estimated further as}
  &\leq
  2 \sum_{k=2}^\infty\frac{\ABS{\tau}^k}{k!}\NORM{H^k\psi_E(\sigma)}.\\
\end{align*}
Since $\psi_E$ is analytic for $U$ in a neighbourhood of $0$, 
also the translates $\psi_E(\sigma)=U_E(\sigma)\psi_E$, 
for $\sigma$ small enough, will be analytic for $U$ in a somewhat smaller
neighbourhood of $0$. This can be seen by noting that the power series
of $U_E(\sigma)$ is term-wise bounded in norm by a convergent one, where
$EHE$ is replaced by $H$, using $\NORM{E}=1$. The composition
of power series in question then amounts to the composition
of analytic functions of $H$ for $\sigma$, $\tau$, small enough.
Therefore the power series on the right hand side of the last inequality
is convergent for $\sigma$, $\tau$ small, 
and defines an analytic function in $\tau$ which is $O(\ABS{\tau}^2)$ as 
$\ABS{\tau}\to0$.
Thus, we finally obtain for small enough $\sigma$, $\tau$ the estimate
\begin{equation}
  \label{eq:UE-EUE-est}
  \NORM{\bigl(U_E(\tau) - EU(\tau)E\bigr)\psi_E(\sigma)}\leq
  \tau^2\cdot C_{\psi_E,\sigma}<\infty.
\end{equation}
Now, from $U_E(t)\psi_E=EU_E(t)E\psi_E$, follows the identity
\begin{equation}
  \label{eq:UE-insert}
  U_E(t)\psi_E=\bigl[EU_E(t/n)E\bigr]^n\psi_E, \quad\text{for all } n,
\end{equation}
by iteration.
Exploiting this, we can rewrite $F_n(t)-U_E(t)$ to yield
\begin{align*}
  \bigl\| & F_n(t)\psi_E - U_E(t)\psi_E\bigr\| =
  \NORM{\bigl[ EU(t/n)E \bigr]^n \psi_E - \bigl[ EU_E(t/n) E\bigr]^n \psi_E}.\\
\intertext{A telescopic estimate shows}
 & \leq 
 \sum_{i=1}^n \NORM{\left\{ 
   \bigl[ EU(t/n)E \bigr]^{n-i}
   \bigl( EU(t/n)E - EU_E(t/n)E \bigr)
   \bigl[ EU_E(t/n)E \bigr]^{i-1}
 \right\}\psi_E}.
\end{align*}
The norm of the vector under the sum is, using~\eqref{eq:UE-insert},
\begin{align*}
  & \NORM{ 
   [ EU(t/n)E ]^{n-i}
   ( EU(t/n)E - EU_E(t/n)E )
   \Psi_E(t(i-1)/n)}\\
\intertext{Using commutativity of $U_E$ with $E$, 
and the invariance of $\Psi_E(\sigma)$ under $E$, we have
$EU_E(t/n)E  \Psi_E(t(i-1)/n)=U_E(t/n) \Psi_E(t(i-1)/n)$,
and use this to rewrite the above expression as}
= & \NORM{ [ EU(t/n)E ]^{n-i} ( EU(t/n)E - U_E(t/n) ) \Psi_E(t(i-1)/n)}\\
\intertext{Now, with $\NORM{\smash{ [ EU(t/n)E ]^{n-i}} }\leq1$ and 
$\NORM{AB\Psi}\leq\NORM{A}\NORM{B\Psi}$, this is bounded by}
\leq &  \NORM{ ( EU(t/n)E - U_E(t/n) ) \Psi_E(t(i-1)/n)}.
\end{align*}
Putting this together, we obtain the estimate
\[
\bigl\|  F_n(t)\Psi_E - U_E(t)\Psi_E\bigr\| \leq
\sum_{i=1}^n \NORM{ ( EU(t/n)E - U_E(t/n) ) \Psi_E(t(i-1)/n)}.
\]
We can now apply~\eqref{eq:UE-EUE-est} to obtain,
for $n>M$ large enough,
\[
 \NORM{ F_n(t)\psi_E - U_E(t)\psi_E} \leq \sum_{i=1}^n 
 \left(\frac{t}{n}\right)^2 \cdot 
 \sup_{\ABS{\sigma}\leq\ABS{t}}C_{\psi_E,\sigma}=
 \frac{t^2C'_{\psi_E,t}}{n},
\]
for some finite $C'_{\psi_E,t}$.
Since $F_n$ converges strongly to $W$ by the AZC, it
follows $W(t)\psi_E=U_E(t)\psi_E$.
The density of the elements $\OA_{E,\tau}\Omega$ 
in $\HS_E$ then shows the claim.
\end{proof}
The explicit form of the generator for the Zeno dynamics
yields an heuristic argument for the equivalence of
the Zeno effects produced by `pulsed' and `continuous'
measurement, respectively. The latter commonly denotes
the simple model for the coupling of the quantum system to a
measurement apparatus that results from
adding to the original Hamiltonian a measurement Hamiltonian 
multiplied by a coupling constant, and letting
the coupling constant tend to infinity~\cite{FP01,FP02,FP02B}. 
The essential point here is that the degrees of freedom
in the Zeno subspace $\HS_E$ become energetically infinitely
separated from those in its orthogonal complement. For
this it suffices to set
\[
H_K\DEF H + K E^\perp,\quad U_K(t)\DEF\ee^{\ii t H_K},
\]
and to consider the limit $K\to\infty$. This can be 
done by applying analytic perturbation theory to
\[
H_\lambda\DEF\lambda H+E^\perp, \quad \text{with }\lambda\DEF K^{-1},
\]
and
\[
U_\lambda(\tau)\DEF\ee^{\ii \tau H_\lambda}=U_K(t), 
\quad \text{with } \tau\DEF Kt=t/\lambda,
\]
around $\lambda=0$. The final result is
\[
\lim_{K\to\infty}U_K(t)\psi=\ee^{\ii t EHE}\psi,
\]
for any vector $\psi\in\HS_E$. 
Details are to be found in~\cite[Section~7]{FP02B}.
Nevertheless, this treatment of `continuous measurement' is certainly
the coarsest possible. To examine more deeply the relationship
between the two manifestations of the Zeno effect, one should
consider more refined models for the interaction of a quantum
with a classical system\eg as in~\cite{BJ93}.

The identification of the Zeno dynamics with the reduced one is in
perfect accordance with the structure of the Zeno subspace $\HS_E$,
showing that the latter is in this case indeed the minimal subspace
to which the Zeno dynamics is restricted (apart from further reducibility
of the Zeno generator). The knowledge about the Zeno generator can be 
used to describe an important class of equilibrium states for 
the Zeno dynamics, namely \textbf{Gibbs equilibria}. For this note
that $U_E$ induces an automorphism group $\widehat{\tau}$ 
of $\OA_E$, as follows from the reasoning of the proof of \refL{auto}.
\refP{reduced} now amounts to the following.
\begin{Lcoro}{explicit-KMS}
  If $(U,E)$ is regular and satisfies~AZC for \OA
then, for every $\beta>0$, 
the set of $(\tau^E,\beta)$-KMS states of $\OA_E$ equals
the set of $(\widehat{\tau}^E,\beta)$-KMS states.
\end{Lcoro}
This result is independent of the representation, 
since the reasoning of \refP{reduced}
can be repeated in any covariant representation.
It applies, in particular,
to the important case of Gibbs states for quantum \textbf{spin systems}.
(for a detailed exposition of these, 
we refer the reader to~\cite[Section~6.2]{b:BR79/81}).
Consider a quantum spin system over the lattice $\XX\DEF\ZZ^d$ with 
interaction $\Phi\colon \XX\supset X\longmapsto \OA_X$. 
The local Hamiltonian of a bounded subset $\Lambda\subset\XX$ is
$H_\Phi(\Lambda)\DEF\sum_{X\subset \Lambda}\Phi(X)$ and
$U_{\Lambda}(t)\DEF\ee^{\ii tH_\Phi(\Lambda)}$ 
is the associated group of unitaries on the finite dimensional, 
local Hilbert space $\HS_\Lambda$.
 The ordinary local Gibbs states over bounded regions $\Lambda\subset\ZZ^d$ are
\[
\omega_\Lambda(A)\DEF
\frac{\Tr_{\HS_\Lambda} 
\bigl(\ee^{-\beta H(\Lambda)}A\bigr)}{%
\Tr_{\HS_\Lambda}\bigl(\ee^{-\beta H(\Lambda)}\bigr)},
\quad\text{for } A\in\OA(\Lambda),
\]
and a candidate for a local Zeno equilibrium over $\Lambda$ is thus
\[
\omega_{E_\Lambda}(A_{E_\Lambda})\DEF
\frac{\Tr_{\HS_{\Lambda}} 
\bigl(\ee^{-\beta E_\Lambda H(\Lambda) E_\Lambda}A_{E_\Lambda}\bigr)}{%
\Tr_{\HS_{\Lambda}}\bigl(\ee^{-\beta E_\Lambda H(\Lambda) E_\Lambda}\bigr)},
\quad\text{for } A_{E_\Lambda}\in\OA(\Lambda)_{E_\Lambda},
\]
if $E_\Lambda\in\OA(\Lambda)$ is some collection of projections, and
where
as before, $\OA(\Lambda)_{E_\Lambda}=E_\Lambda \OA(\Lambda) E_\Lambda$.
Here it is safe to take the trace over the full local space $\HS_\Lambda$, since 
\[
\omega_{E_\Lambda}(A B_{E_\Lambda} C)=
\omega_{E_\Lambda}(A_{E_\Lambda} B_{E_\Lambda} C_{E_\Lambda}),
\]
for $A$, $B$, $C\in\OA(\Lambda)$, as follows easily from
$E\ee^{-\beta E_\Lambda H(\Lambda)E_\Lambda}=
\ee^{-\beta E_\Lambda H(\Lambda)E_\Lambda}E=
\ee^{-\beta E_\Lambda H(\Lambda)E_\Lambda}$ 
and the invariance of the trace under cyclic permutations.

Assume that the local dynamics $\tau_t^\Lambda$ generated by $H(\Lambda)$
converges uniformly  to an automorphism group $\tau$ of \OA. Then
we know~\cite[Proposition~6.2.15]{b:BR79/81}, that every thermodynamic
limit point of the ordinary local Gibbs states,
that is, a weak* limit of a net of extensions $\omega_\Lambda^G$ of 
$\omega_\Lambda$ to \OA, is a $(\tau,\beta)$-KMS state over \OA.
As a direct consequence of these considerations 
and~\refC{explicit-KMS}, 
we obtain those equilibrium states for the Zeno dynamics which are limits
of local Gibbs states.
\begin{Lcoro}{gloGibbs}
Let $\beta>0$. Let $\Lambda_\alpha\to\infty$ be such that the local dynamics
converges uniformly to the global dynamics $\tau$, 
and the net of local Gibbs states $\omega_{\Lambda_\alpha}$ 
has a thermodynamic limit point $\omega$. 
Let $U$ be the unitary group representing $\tau$ 
in the GNS representation of $\omega$.
If a sequence of projections $E_{\Lambda_\alpha}\in\OA(\Lambda_\alpha)$
converges in norm to a projection $E$ in \OA
such that $(U,E)$ is regular and satisfies~(AZC),
then $\omega_E(A_E)\DEF\lim_\alpha\omega_{E_{\Lambda_\alpha}}^G(A_E)$ 
defines  a $(\tau^E,\beta)$-KMS state on $\OA_E$.
\end{Lcoro}
\begin{proof}
  The local Gibbs states $\omega_{E_{\Lambda_\alpha}}$ are the unique $\beta$-KMS states
on the finite-dimensional algebras $\OA_{\Lambda_\alpha}$ for the 
reduced dynamics $\widehat{\tau}^{E_{\Lambda_\alpha}}$.
If $\{E_{\Lambda_\alpha}\}$ converges uniformly, these
local Gibbs state possess $\omega_E$ as a weak-* limit,
which is a KMS state on $\OA_E$
for the reduced dynamics $\widehat{\tau}^E$
associated with $\tau$.
Then, by ~\refC{explicit-KMS}, $\omega_E$ is also a $(\tau^E,\beta)$-KMS state.
\end{proof}
We finally note the \textit{dynamical} manifestation of the Zeno paradox
in quantum statistical mechanics. Assume that the difference between
the Zeno generator $EHE$ and the original one $H$ is entire analytic
for the Zeno dynamics $\tau^E$. Then, the
original dynamics is a local perturbation of the Zeno dynamics,
and the general results about the return to 
equilibrium~\cite[Theorem~2]{ROB73}, which have been described in
Section~\ref{sec:AZC-app},
imply that the system starting in a global equilibrium state
for the dynamics defined by $H$ will spontaneously evolve
toward a KMS state for the Zeno dynamics.
\begin{Lcoro}{Zeno-equi}
Let $(U,E)$ be regular and satisfy~AZC for \OA.
  Let $\omega|_{\OA_E}$ be the restriction of a $(\tau,\beta)$-KMS state
of \OA to $\OA_E$. Assume that $(\OA_E,\tau^E)$ is asymptotically
Abelian, and that $H-EHE$ is entire analytic for $\tau^E$.
Then, every weak* limit point for $t\to±\infty$ 
of $\tau^E_t\omega|_{\OA_E}$ 
is a $(\tau^E,\beta)$-KMS state.
\end{Lcoro}
\subsection{A Condition on the Zeno Generator}
\label{sec:EI}
We return from the operator algebraic framework of quantum statistical
mechanics to general, unitary groups on a Hilbert space, to present
what appears as the most advanced, functional analytical condition for the existence 
of Zeno dynamics so far. 
Although this very recent result~\cite{EI03} proposes,
like all other ones in this Section, a \textit{sufficient} condition,
it seems to be the sharpest, general characterisation of Zeno dynamics 
presently available (the necessary and sufficient conditions below in
Section~\ref{sec:EDF} are less general and require detailed, additional
information about Hamiltonian and projection). It also clarifies the exact
form of the Zeno generator and determines the Zeno subspace completely.

Let $H$ be a nonnegative self-adjoint
operator on a separable Hilbert space \HS, and $E$ an orthogonal projection
on \HS. The quadratic form 
\[
\phi\longmapsto\NORM{\smash{H^{1/2}E}\phi}, \quad\text{with form domain }
D(H^{1/2}E),
\]
has a self-adjoint operator
\[
H_E\DEF (H^{1/2}E)^\ast(H^{1/2}E)
\] 
associated with it. The necessary condition is now formulated in terms of
$H_E$. 
\begin{theo}[{\cite[Corollary~2.2]{EI03}}]\labelT{EI}
  If $H_E$ as defined above is densely defined on \HS, then holds
\[
\slim_{n\to\infty}\Bigl[ E \ee^{\ii\frac{t}{n}H} \Bigr]^n =
\slim_{n\to\infty}\Bigl[  \ee^{\ii\frac{t}{n}H}E \Bigr]^n =
\slim_{n\to\infty}\Bigl[ E \ee^{\ii\frac{t}{n}H}E \Bigr]^n =
\ee^{\ii t H_E}E,
\]
uniformly in $t$ on every compact interval in \RR.
\end{theo}
Note that the result in~\cite{EI03} concerns a yet more general case when
the evolution is interrupted by \textit{different} projections
$E(\theta t/n)$ at the times $t/n$, which render a strongly continuous,
non-increasing function $E(t)$, with the initial condition
$\slim_{t\to 0}E(t)=E$. This generality is not needed for our 
present discussion.
The proof of the main theorem in~\cite{EI03}, which entails also
the result above, is a clever combination of the known product formulae
of Chernoff~\cite{CHE68,b:CHE74} with the results of Kato~\cite{KAT78}
and Ichinose~\cite{ICH80},
and we will not go into details. 

So, the Zeno dynamics exists, if only its generator $H_E$ is densely
defined --- a very weak, yet nontrivial, condition, see the counterexample
in~\cite{EI03}. It is also not completely dissimilar to the regularity
condition that allowed us to identify the Zeno generator in the more
benign case treated in \refP{reduced}. In general, $H_E$ is not densely
defined but is a self-adjoint operator on the closed subspace of \HS,
determined as the closure of the form domain $D(H^{1/2}E)$, 
\[
\widehat{\HS}_E\DEF D(H^{1/2}E)\subset E\HS,
\]
which is now the relevant Zeno subspace, in general larger than the
Zeno subspace $\HS_E$ in the algebraic context above.
The requirement that $H_E$ be densely defined amounts to saying that
the form domain $D(H^{1/2}E)$ is dense in \HS. 
Furthermore, $H_E$ differs from the reduced generator $EHE$,
which is not necessarily closed, since $EH$ does not need to be closed
though $HE$ is. In fact, $H_E$ is generally a self-adjoint extension
of $EHE$.
\section{Physical Considerations and the anti--Zeno effect}
\label{sec:phys}
\subsection{Geometry of the Hilbert Space}
\label{sec:geometry}
To show a further, and arguably more fundamental facet of the reasons
leading to Zeno effect and paradox, we briefly describe the role that
is played in that piece by the geometry of the Hilbert space. We follow
mainly the clear account of~\cite{PL98}, with some borrowings from~\cite{AA90}
and~\cite{PAT95}.
We assume that some generalised quantum evolution --- without any 
supposition about linearity, group structure, or unitarity ---
acts smoothly in the vicinity of a point $\psi(0)$ in a separable Hilbert space \HS
(in the natural topology of \HS). Let the system be prepared in the 
initial state $\psi(0)$, which shall be an eigenvector of the relevant
observable $\mathcal{O}$. In turn, $\mathcal{O}$ is assumed to possess a complete set of
eigenvalues $\{\mathcal{O}_n\}$ and eigenfunctions $\{\psi_n\}$. The \textbf{survival
probability} of the initial state at time $t$ is then
\[
  \mathcal{P}(t)\DEF \ABS{\IPROD{\frac{\psi(0)}{ \NORM{\psi(0)}}}{
\frac{\psi(t)}{\NORM{\psi(t)}}
}}^2,
\]
since the evolving vector need not be normalised at later times, 
and can change its norm during the evolution.
We introduce the vector $\chi(t)\DEF\psi(t)/\NORM{\psi(t)}$,
which is always normalised. Taylor expansion yields
the asymptotics
\[
\chi(\tau)=\chi(0)+\tau\dot{\chi}(0)+\frac{\tau^2}{2}\ddot{\chi}(0)+O(\tau^3)
\quad(\tau\to 0),
\]
and hence for the survival amplitude
\[
\mathcal{A}(\tau)=\IPROD{\chi(0)}{\chi(\tau)}
=1+\tau\IPROD{\chi(0)}{\dot{\chi}(0)}+
\frac{\tau^2}{2}\IPROD{\chi(0)}{\ddot{\chi}(0)}+O(\tau^3)
\quad(\tau\to 0).
\]
But since $\chi$ is kept normalised, the easy calculation
\[
0=\left.\frac{\dd\NORM{\chi(t)}^2}{\dd t}\right|_{t=0}=
\IPROD{\chi(0)}{\dot{\chi}(0)}+\IPROD{\dot{\chi}(0)}{\chi(0)}=
2\RE\IPROD{\chi(0)}{\dot{\chi}(0)}
\]
shows that $\IPROD{\chi(0)}{\dot{\chi}(0)}$ is purely imaginary.
>From this, we obtain for the survival probability $\mathcal{P}(\tau)=
\ABS{\mathcal{A}(\tau)}^2$ the expression
\begin{alignat*}{2}
  \mathcal{P}(\tau) &=
  1 &+ & \tau  \Bigl[ \IPROD{\chi(0)}{\dot{\chi}(0)}+\IPROD{\dot{\chi}(0)}{\chi(0)} \Bigr]\\
& & + & \tau^2 \left[ \IPROD{\chi(0)}{\dot{\chi}(0)}\IPROD{\dot{\chi}(0)}{\chi(0)} 
                     +\frac{1}{2}\bigl\{\IPROD{\chi(0)}{\ddot{\chi}(0)}+
                         \IPROD{\ddot{\chi}(0)}{\chi(0)}\bigr\} \right]+O(\tau^3)\\
& = 1 &+ & 2 \tau \RE\IPROD{\chi(0)}{\dot{\chi}(0)}
 + \tau^2 \left[ \ABS{\IPROD{\chi(0)}{\dot{\chi}(0)}}^2+
               \RE\IPROD{\ddot{\chi}(0)}{\chi(0)}\right]+O(\tau^3)\\
& = 1 & + & \tau^2 \left[ \left(\IM\IPROD{\chi(0)}{\dot{\chi}(0)}\right)^2+
               \RE\IPROD{\chi(0)}{\ddot{\chi}(0)}\right]+O(\tau^3)
\quad(\tau\to 0).
\end{alignat*}
In a similar manner as above we see
\[
0=\left.\frac{\dd^2\NORM{\chi(t)}^2}{\dd t^2}\right|_{t=0}=
2\left[ \IPROD{\dot\chi(0)}{\dot\chi(0)}+\RE\IPROD{\chi(0)}{\ddot\chi(0)}\right],
\]
that is, $\RE\IPROD{\chi(0)}{\ddot\chi(0)}=-\IPROD{\dot\chi(0)}{\dot\chi(0)}$,
allowing us to write
\[
\mathcal{P}(\tau)= 1-\tau^2 k+O(\tau^3)
\ (\tau\to 0) \quad\text{with }
k=\IPROD{\dot\chi(0)}{\dot\chi(0)}-\left(\ii \IPROD{\chi(0)}{\dot\chi(0)}\right)^2.
\]
Let us assume for the moment that $k$ is nonnegative. Then, if $N$ measurements
of $\mathcal{O}$ are performed at consecutive times $\tau_i=i t/N$, $i=1,\ldots,N$,
the final survival probability of the initial state at $\tau_N=t$ is
\[
\mathcal{P}_N(t)=
\left[\mathcal{P}(t/N)\right]^N=
\left[ 1-\frac{t^2k}{N^2} \right]^N+O(N^{-3N})
\sim \ee^{kt^2/N^2}\quad\text{for $N$ large},
\]
where, of course, the collapse \textit{viz.} projection postulate
has implicitly been applied by decomposing $\mathcal{O}$ into its 
eigenprojections. In the limit of infinitely frequent measurement,
we then recover the Zeno paradox.
\[
\lim_{N\to\infty}\mathcal{P}_N(t)=1.
\]
This means that, apart from the crucial assumption of non-negativity of the
constant $k$ to which we will come shortly, the Zeno effect is essentially
a consequence of the \textit{projective} nature of the quantum formalism.
For the calculations above use at the decisive steps the postulate that
probabilities are calculated from \textit{normalised} vectors by taking
absolute squares of their inner products\ie from unit rays in Hilbert space 
or \textit{states}. The Zeno effect 
therefore does not hinge on the particularities of the evolution in question,
but rather on the use of the Hilbert space formalism and the probability 
interpretation --- indeed, two fundaments of quantum theory.

Let us come back to the constant $k$ and the question of its non-negativity.
In the case of unitary evolutions generated by a Hamiltonian $H$, it
can be identified as the expected variance of $H$ in the initial state
\[
(\Delta H)^2=\IPROD{\psi}{H^2\psi}-\IPROD{\psi}{H\psi}^2,
\]
where $\psi=\chi(0)$ is the normalised, initial state. 
This holds if the first
moment of $H$ in the state $\psi$ is finite, as follows
from the asymptotic expansion of the survival amplitude 
at the end of Section~\ref{sec:AZC}. This physical quantity is always
nonnegative and we obtain yet another proof that the Zeno paradox
emerges in this case. We want to show that also in the general
case $k$ allows for a physical interpretation which makes
its non-negativity very credible. In the general case at hand,
we first have to take the correct perspective by looking at
the projective state space $\PP=\HS^\ast/U(1)$, of the Hilbert space \HS, 
where $\HS^\ast=\{\psi\in\HS\mid\NORM{\psi}=1\}$ is the unit sphere of
\HS. The set \PP of unit rays of \HS can be equipped with a natural 
metric which arises from the inner product of vector representatives by
\[
s^2=4\left(1-\ABS{\IPROD{\frac{\psi}{\NORM{\psi}}}{\frac{\phi}{\NORM{\phi}}}}^2\right).
\]
It is a measure of the distance between points in \PP and satisfies
the usual metric axioms there.
The metric is then given in infinitesimal form by its line element
\[
\dd s^2 =
4\left[ 
\IPROD{\dot\chi(t)}{\dot\chi(t)}
-\left( \ii \IPROD{\chi(t)}{\ddot\chi(t)}\right)^2
\right]
\dd t^2,
\]
where as before $\chi(t)\DEF\psi(t)/\NORM{\psi(t)}$.
This metric, constructed in~\cite{PAT95}, 
is a generalisation of the Fubini-Study 
metric~\cite{AA90} and reduces to it in the case of linear, unitary 
evolutions. Denoting by $v(t)=\dot{s}(t)$ the reparametrisation invariant
speed at which a point in \PP travels under the evolution, we see
immediately that
\[
k=v(0)^2/4
\]
is the square of the initial speed, a quantity whose non-negativity is 
guaranteed. Indeed at this point, the Zeno paradox appears as unavoidable, 
at least for rank-one projections onto initial states which evolve
smoothly. The only conceivable possibility for not ending up in the Zeno
regime remains when the asymptotics used in the above reasoning becomes 
unreliable hence, in the Hamiltonian context, for states with energetic
singularities.
\subsection{The Zeno -- Anti-Zeno Transition}
\label{sec:Z-AZ}
In contrast to the conclusions of the last section, 
 we want to demonstrate that quantum evolution can not only be 
impeded by frequent measurement, but that it can also be
\textit{accelerated}, a surprising phenomenon which has aptly
been termed \textbf{inverse Zeno}, \textbf{anti--Zeno}, or
\textbf{Heraclitus} effect (due to Heraclitus' reply 
``everything flows'' to 
Zeno's argument). 
We follow~\cite{FNP01,FP01,FP01A}, and first rephrase
the Zeno effect in terms of \textbf{decay rates}.
It is well known that for sufficiently long times, an
unstable quantum system shows exponential decay\ie
the survival probability of the initial state approaches
\[
\mathcal{P}(t)\sim\mathcal{Z}\ee^{-\gamma_0 t}
\quad\text{for large $t$},
\]
according to its natural decay rate $\gamma_0$, and
where the positive constant $\mathcal{Z}$ can be identified in
field theoretical models as the wave function renormalisation
constant. On the other hand, we already learnt about
the quadratic behaviour of $\mathcal{P}$ at short times
(in the cases where the Zeno effect persists)
\[
\mathcal{P}(\tau)\sim 1-\tau^2/\tau_{\text{Z}}^2
\quad(\tau\to0),
\]
where $\tau_{\text{Z}}^{-2}=(\Delta H)^2$ is called the \textbf{Zeno time}.
Thus, the survival probability will generally interpolate between 
these two regimes, which can be expressed, 
using an \textbf{effective decay rate}
$\gamma_{\text{eff}}(\tau)$, as
\[
\mathcal{P}(\tau)=\ee^{-\gamma_{\text{eff}}\tau},
\quad\text{with }
\gamma_{\text{eff}}(\tau)=-\frac{1}{\tau}\ln\mathcal{P}(\tau).
\]
The effective decay rate interpolates between the quadratic short time 
and exponential regimes as
\[
\gamma_{\text{eff}}(\tau)\sim
\begin{cases}
  \tau^2/\tau_{\text{Z}}^2 & (\tau\to 0),\\
  \gamma_0 & (\tau\to\infty).
\end{cases}
\]
Now, if there exists a time $\tau^\ast$ with
\[
\gamma_{\text{eff}}(\tau^\ast)=\gamma_0
\]
then measurements of the undecayed state 
performed at intervals $\tau^\ast$ let the system decay at
its natural rate $\gamma_0$, that is, as if no measurements
were performed. In turn, if such a (unique) 
$\tau^\ast$ exists it means
that for shorter measurement intervals the decay will be 
inhibited, since $\gamma_{\text{eff}}$ is smaller than
$\gamma_0$ in that region --- this is the regime of the quantum Zeno 
effect. Yet, for measurement intervals $\tau>\tau^\ast$ one
then has generically $\gamma_{\text{eff}}>\gamma_0$,
corresponding to an accelerated decay --- the anti--Zeno regime.

A sufficient, and physically meaningful, condition for the
existence of at least one $\tau^\ast$ is $\mathcal{Z}<1$. For then
the graph of the survival probability starts out \textit{above}
the exponential $\ee^{-\gamma_0 \tau}$ (due to the quadratic short 
time behaviour), but ends up approximating the renormalised
exponential $\mathcal{Z}\ee^{-\gamma_0 \tau}<\ee^{-\gamma_0 \tau}$,
and thus must have an intersection with it. The appearance of the
anti--Zeno effect has in fact been experimentally confirmed~\cite{FGMR01}, 
see also~\cite{BR01,KS02,GUR02} for further theoretical considerations.
\subsection{Asymptotics of State Energy Distribution}
\label{sec:EDF}
We finally come to the very recent results of Atmanspacher, Ehm and 
Gneiting~\cite{AEG03}, who show that a sharp characterisation 
of the transition between Zeno and anti--Zeno effect can be given 
using the energy distribution of the decaying
state. More precisely, the relevant information is encoded
in the asymptotic decay of the cumulative energy density function
of the initial state. The results 
are formulated in the framework
of probability theory, and we briefly introduce the necessary notions.
Consider independent random variables $X_1, X_2,\ldots$ distributed
according to a common law $\Pr(X_k<x)=F(x)$, where $F$ is some probability 
distribution on \RR\ie an non-decreasing, left continuous function with
$\lim_{x\to-\infty}=0$, and $\lim_{x\to\infty}=1$.
Its \textbf{characteristic function} is given by the Fourier transform
\[
\varphi(t)\DEF\int_{-\infty}^\infty \ee^{-\ii t x}\dd F(x),
\]
while its decay at infinity is captured in the quantity
\[
\delta_F(x)=x\Pr\bigl(\ABS{X_k}>x\bigr)=x\bigl(F(-x)+1-F(x)\bigr),
\]
where the last equality holds at all points of continuity of $F$.
Let us relate these notions to their physical counterparts.
If we interpret $F(x)$ as the cumulative energy distribution of an initial, decaying
quantum state $\psi$\ie as the probability to measure energies of
absolute value larger than $x$ in this state, then $\varphi$ corresponds to
the time evolution of this state, more precisely to the survival
amplitude $\mathcal{A}(t)$. Namely, $\mathcal{A}(t)$ can be expressed as
\[
\mathcal{A}(t) = \int_{-\infty}^\infty \ee^{-\ii tE} \ABS{\lambda(E)}^2\dd E,
\]
(here we changed a sign in contrast to our previous notation)
where $\dd F$ is in fact identified as the absolute square of
the energy density function $\lambda(E)$ of $\psi$, in its decomposition
into 
energetic components.
Note aside, that in $F$ also negative energies are allowed, and that the results
are insensitive to that, which is another example for the fact
that semiboundedness is not required for the Zeno effect.
Thus, the independent probability variables $X_k$ are nothing but
the outcomes of energy measurements of the system. We are now ready to
state the first main result.
\begin{theo}[{\cite[Theorem~1]{AEG03}}]\label{theo:AEGZ}
  Equivalent are

\noindent i) $\lim\limits_{n\to\infty}\ABS{\varphi(t/n)}^{2n}=1$
for all $t\in \RR$.

\noindent ii) $\lim\limits_{x\to\infty}\delta_F(x)=0$.

\noindent iii) For all $\EPS>0$ holds
\[
\lim\limits_{n\to\infty}\inf_{\mu\in\RR}
\Pr\left( \ABS{\frac{X_1+\cdots+X_n}{n}-\mu}>\EPS\right)=0,
\]
that is, there exists a sequence of real numbers $\mu_n$ such that
the distribution of the averages $(X_1+\cdots+X_n)/n-\mu_n$ converges
weakly to the Dirac measure concentrated at zero.
\end{theo}
\begin{proof}
  The weak law of large numbers~\cite{KAL02} ensures equivalence of ii) 
and iii), thus it remains to show the equivalence of i) and ii).
The function $\gamma(t)=\ABS{\varphi(t)}^2$ is the characteristic function
of the difference $Y=X'-X''$ of independent random variables $X'$, $X''$, 
each with distribution $F$. If $G$ denotes the distribution function of $Y$ 
then $\gamma(t/n)^n$ is the characteristic function of the mean 
$(Y_1+\cdots Y_N)/n$ of $n$ independent random variables with
distribution $G$. Then
\[
\lim\limits_{n\to\infty}\ABS{\varphi(t/n)}^{2n}=
\lim_{n_\to\infty}\gamma(t/n)^n=1\quad\text{for all }t\in\RR
\]
holds if and only if $(Y_1+\cdots Y_N)/n$ converge to zero in distribution, 
which in turn is equivalent to 
\[
\lim_{x\to\infty}\delta_G(x)=0,
\]
since $G$ is symmetric. But due to the symmetrisation inequalities
\[
\exists a\colon \forall x>0\colon\quad
\frac{1}{2}\Pr\bigl(\ABS{X'}>x+a\bigr) \leq
\Pr\bigl(\ABS{X'-X''}>x\bigr)\leq
2\Pr\bigl(\ABS{X'}>x/2\bigr),
\]
for probability distributions~\cite[p.~149]{b:FEL71}, 
the latter is equivalent to ii).
\end{proof}
Very remarkably, a similar characterisation of the anti--Zeno effect
was also achieved in~\cite{AEG03}. Here, for the first time, the
anti--Zeno effect is considered in the infinitely frequent measurement
limit\ie the \textbf{anti--Zeno paradox} is treated, and shown to
lead to a spontaneous decay, that is, to vanishing survival probability
at arbitrary small times. This is the last result we reproduce, and we
omit the proof, which is again based on the law of large numbers, 
but is a bit more involved.

We need a mild regularity condition on $F$, and say that $F$ is
\textbf{straight} if either $\sup_{x>0}\delta_F(x)<\infty$
or $\lim_{x\to\infty}\delta_F(x)=\infty$. 
This excludes cases where $\limsup_{x\to\infty}\delta_F(x)=\infty$
while $\lim_{x\to\infty}\delta_F(x)$ does not exist,
roughly corresponding to energy spectra with a sequence of gaps of 
increasing size.
\begin{theo}[{\cite[Theorem~2]{AEG03}}]\label{theo:AEGaZ}
  If $F$ is straight then the following conditions are equivalent

\noindent i) $\lim\limits_{n\to\infty}\ABS{\varphi(t/n)}^{2n}=0$ in
measure\ie for all $T>0$ and $\EPS>0$, the Lebesgue measure of the set
of all $\ABS{t}<T$ with $\ABS{\varphi(t/n)}^{2n}>\EPS$ converges to zero.

\noindent ii) $\lim\limits_{x\to\infty}\delta_F(x)=\infty$.

\noindent iii) For all $c>0$
\[
\lim\limits_{n\to\infty}\sup_{\mu\in\RR}
\Pr\left( \ABS{\frac{X_1+\cdots+X_n}{n}-\mu}\leq c\right)=0,
\]
that is, the distribution of $(X_1+\cdots+X_n)/n$ spreads out
over \RR as $n\to\infty$.
\end{theo}
The crucial observation that enables the application of the weak 
law of large numbers in the proofs of both theorems,
is the reinterpretation of the iterated survival amplitude
$\mathcal{A}(t/n)^n$. It is now seen as the characteristic function of
the mean value of $n$ energy measurements, carried out on an
ensemble of quantum systems prepared in the state $\psi$. 

The physical interpretation of the results is
that the critical transition between Zeno and anti--Zeno effect occurs
roughly at state energy distributions for which the probability
of measuring energies larger than $E$ decays as $1/E$. More 
precisely, for instance condition~ii) in Theorem~\ref{theo:AEGZ} 
corresponds to $o(1/E)$. The characterisation of the anti--Zeno
effect, or rather the anti--Zeno paradox of Theorem~\ref{theo:AEGaZ}
is somewhat at  variance with the heuristic presentation of the
Zeno -- anti--Zeno transition in the last section. The latter
prevails only for \textit{nonzero} measurement intervals, while
here we are again concerned with the infinitely frequent measurement
limit. 

The distinction between Zeno and anti--Zeno regime shown by the above two
theorems is much finer than that of Luo \textit{ et al.}~\cite{LWZ02}
who showed that finiteness of the first absolute moment 
\[
\int_{-\infty}^\infty \ABS{E} \ABS{\lambda(E)}^2\dd E,
\]
of the energy density $\lambda(E)$ of the initial state,
is sufficient for the Zeno paradox. 
%
\section{Concluding Remarks}
\label{sec:conc}
The Zeno effect was once dismissed as a curious paradox
that could only emerge in thought experiments based on wrong
concepts of quantum theory and physical reality. Yet the
phenomenon has lived through a renaissance. By now the effect
appears as one of the most generic ones in quantum theory,
and as extremely robust with respect to special formulations, 
ancillary conditions, and the wide range of specific models
that has by now been considered. Even realistic possibilities for 
its application are under serious consideration~\cite{FHKPR02}. 

The main line of thought of the present survey of the mathematics
of the Zeno effect is, in retrospect, the challenge to delineate
its domain of validity.
The general picture that emerges is that the prevalence of the
Zeno effect can only be broken in exceptional cases, for which also 
the nature of the mathematical counterexamples constructed so 
far~\cite{AU97,MS01,MAS02} is an indication. In view of
Section~\ref{sec:geometry}, what is required for leaving the
Zeno regime, at arbitrary short times, 
is a certain amount of non-analyticity of the evolution
starting in the initial state.
Energetic singularities of this state, respectively, a 
state in the Zeno subspace seem to be a fundamental prerequisite
for this, a view which is 
corroborated
by the conditions
for the anti--Zeno paradox of Theorem~\ref{theo:AEGaZ}.
On one hand such singularities \textit{are} in fact present
in field theoretical descriptions of decay processes, which
is related to the discovery of the possibility of a regime governed
by the anti--Zeno effect as described in Section~\ref{sec:Z-AZ}, 
following~~\cite{FNP01,FP01,FP01A}, cf.\ also~\cite{FP02A}. 
On the other hand, whether there exist physical systems
which could exhibit the anti--Zeno paradox according to
the conditions of Theorem~\ref{theo:AEGaZ}, is an open
question, and might even be considered doubtful.

>From a mathematical viewpoint, it would be most important
to obtain sharp conditions for the Zeno paradox and
the anti--Zeno paradox. For the former,
the results of Section~\ref{sec:EI} are the
best as yet, in a general operator theoretical framework, while 
for both Zeno and anti--Zeno paradox, the results of Section~\ref{sec:EDF} 
are the most advanced for rank one projections.
They are also closest in spirit to our heuristic reasoning
of Section~\ref{sec:conc1}, which also led us to conjecture that
the asymptotic growth of the energy density, set into proper relation
to the projection, should be indicative for the Zeno effect.
Yet, the probabilistic argument used in this section are of a quite 
different quality than the Payley--Wiener type, and 
complex analytic argument
that was coarsely conceived in Section~\ref{sec:conc1}. The interesting
question remains whether it is possible to find 
corresponding results
in the latter framework, which would also generalise the results
of Section~\ref{sec:EDF} to projections of infinite rank.

Another subject worth further theoretical work is the identification
of the Zeno dynamics and the Zeno subspace. It appears quite generally
to be an ordinary quantum dynamics, which is confined, in the
Zeno limit, to the Zeno subspace by additional boundary conditions.
The prime example is the projection operator given by multiplication with
the characteristic function of an interval of the real line, leading to
Dirichlet boundary conditions~\cite{FPSS01,EI03}. This has been followed
by the identification of the Zeno generator in Section~\ref{sec:gen-equi}
as $EHE$, possible under certain technical conditions, and the finer
characterisation of~\cite{EI03}. In~\cite{AUS03} we have considered
models of quantum spin systems, which exhibit the underlying mechanism,
and point out possible manifestations of the Zeno effect in mesoscopic,
or even macroscopic systems.
Yet a general formulation and classification of the emerging boundary
conditions would be desirable, and seems conceivable in the operator 
algebraic framework. Such general results would probably also be applicable in
the 
context of algebraic quantum field 
theory~\cite{b:HAA92}, where the Zeno effect has, as of yet, 
not received much attention.
\pagebreak
\newcommand{\noopsort}[1]{} \newcommand{\singleletter}[1]{#1}
\providecommand{\bysame}{\leavevmode\hbox to3em{\hrulefill}\thinspace}
\providecommand{\MR}{\relax Math.~Rev.~}

\end{document}